\documentclass{aa}  
\usepackage{graphicx}
\usepackage{txfonts}
\usepackage{multirow}
\usepackage{xcolor}
\usepackage{hyperref}
\hypersetup{
    colorlinks,
    linkcolor={blue!60!black},
    citecolor={blue!60!black},
    urlcolor={blue!60!black}
}

\newcommand\B{\langle B \rangle}
\newcommand\teff{T_{\rm eff}}
\newcommand{\kms}{\rm km\,s^{-1}}

\begin{document}

      \title{A fast method for deriving relative small-scale magnetic field variations from high-resolution spectroscopy}
      \titlerunning{A fast method for deriving relative small-scale magnetic field variations}

   \author{Paul I. Cristofari
          \inst{1}
          \and
          Steven H. Saar
          \inst{2}
          \and Aline A. Vidotto
          \inst{1}
          \and
          Stefano Bellotti
          \inst{1}
          }

   \institute{Leiden Observatory, Leiden University, PO Box 9513, 2300 RA Leiden, The Netherlands\\
              \email{cristofari@leidenuniv.nl}
         \and
             Center for Astrophysics $\vert$ Havard \& Smithsonian, 60 Garden Street, Cambridge, MA 02138, USA
             }

   \date{Received Month DD, YYYY; accepted Month DD, YYYY}

 
  \abstract
   {Setting observational constraints on stellar magnetic fields is essential for both stellar and planetary physics. They play a key role in the formation and evolution of stars and planets, and they are responsible for spurious signals in radial velocity curves that impact the detection and characterization of exoplanets. Recent observations have revealed the diversity and evolution of large-scale magnetic fields in low-mass stars. However, these large-scale fields only account for a small fraction of the observed unsigned magnetic flux. 
   The other crucial stellar magnetism information originates from (spatially) small-scale magnetic fields, which account for most of the surface magnetic flux and exhibit a clear temporal evolution on timescales of many years.
        }
   {With this work, we aim to develop new fast techniques to extract small-scale magnetic field estimates from time series of observed high-resolution spectra. One objective is to develop tools that will enable the community to take full advantage of the upcoming monitoring surveys carried out with various high-resolution spectrometers. Our ultimate goal is to study the temporal evolution of small-scale magnetic fields and offer insights into the magnetic properties of low-mass stars and their magnetic cycles.}
   {We implemented a process to capture relative pixel variations caused by changes in magnetic field strengths, relying on synthetic spectra computed with \texttt{ZeeTurbo}. This approach provides extremely fast and reliable estimates of relative magnetic field strength variations from series of high-resolution spectra, mitigating the impact of systematics between models and observations. We assessed the performance of the proposed method through its application to simulated data and publicly available observed spectra recorded with SPIRou, Narval, and ESPaDOnS. In addition, we implemented a model-driven process to derive relative temperature variations and we explored the influence magnetic fields have on these measurements.}
   {Our results are in excellent agreement with the magnetic field estimates previously obtained from spectra recorded with SPIRou. This method provides robust constraints on the structure of the magnetic field variations and proves to be relatively insensitive to small changes in the assumed atmospheric parameters and broadening. 
        We find that magnetic field variations have the potential of introducing biases in relative temperature estimates. This is particularly relevant in the case of the Narval/ESPaDOnS spectral domains, which contain a large number of magnetically sensitive transitions and where contrast is more important.
        Our application to archival data provides new constraints on the evolution of small-scale magnetic fields and underscores the potential of the proposed method for  analyzing data in the context of large observation programs.}
   {By reducing the problem to a set of linear equations, our method offers extremely fast results, making it viable for integration in future pipelines developed for large spectroscopic surveys. These estimates will provide much needed information to correct radial velocity curves and constrain dynamo processes.}
   

   \maketitle
%

	\nolinenumbers

\section{Introduction}

    Sun-like stars and M dwarfs are known to host magnetic fields~\citep[e.g.,][]{saar-1985, johns-krull-1996, kochukhov-2021, reiners-2022} that have consequences on their polarized and unpolarized spectra. These fields are thought to be responsible for activity phenomena, introducing spurious signal in radial velocity (RV) measurements~\citep{dumusque-2021, bellotti-2022} and consequently impacting exoplanet detection and characterization. Magnetic fields also play a crucial role in stellar formation and evolution processes~\citep{donati-2009}, leading to angular momentum loss~\citep[e.g.,][]{skumanich-1972, vidotto-2014} and surface inhomogeneities (e.g., spots, plages, and faculae). 

    M dwarfs are the most numerous stars in the solar neighborhood~\citep{reyle-2021} and they are considered prime targets for the search of habitable exoplanets. The precise and accurate characterization of these stars is essential to extract reliable estimates on the masses and radii of the exoplanets orbiting them~\citep{bonfils-2013, dressing-2015, gaidos-2016}. The masses of M dwarfs range from 0.08 to 0.57\,$\rm M_\odot$~\citep{pecaut-2013}: stars with masses below 0.35\,$\rm M_\odot$ are believed to be fully convective~\citep{chabrier-1997}, while M dwarfs with masses above 0.35\,$\rm M_\odot$ have radiative interiors and an outer convective envelope separated by a transition region known as the tachocline~\citep{spiegel-1992}.
    The tachocline is thought to play a key role in the 22-year magnetic cycle observed in the Sun~\citep[see][]{parker-1955, charbonneau-2010}. In fully convective M~dwarfs, dynamo processes must be revisited and alternative scenarios have been proposed to generate large-scale fields without the need for a tachocline~\citep{chabrier-2006, yadav-2015}. 

        The last decade has been marked by the rapid development of high-resolution spectrometers, including several near-infrared instruments such as CARMENES~\citep{quirrenbach-2014}, CRIRES+~\citep{dorn-2023}, and SPIRou~\citep{donati-2020}, motivated in part by the search for exoplanets around M~dwarfs. The spectropolarimetric capabilities of some of these instruments have led to numerous studies focused on large-scale stellar magnetic fields~\citep[e.g.,][]{kochukhov-2020, finociety-2023, donati-2023b,bellotti-2024}, while unpolarized spectra have enabled the community to characterize the small-scale magnetic fields of low-mass stars~\citep[e.g.,][]{shulyak-2017, reiners-2022, cristofari-2023, cristofari-2023b}. Combined efforts to characterize both large-scale and small-scale magnetic fields have provided a complete picture of cool stars magnetism~\citep[e.g.,][]{kochukhov-2017, donati-2023, donati-2025, bellotti-2025}.
    Recent works have reported efforts to characterize the variation of small-scale magnetic fields throughout observation campaigns lasting several years~\citep{donati-2023, cristofari-2025b} and revealing rotational modulation and long-term fluctuations. Small-scale magnetic fields have been proposed as an excellent proxy to correct for activity jitters in radial velocity curves~\citep{haywood-2016, haywood-2022}. The extraction of small-scale field fluctuations from the high-resolution spectra, particularly from the spectra used to obtain radial velocity curves, is therefore  timely.
    
    With this work, we propose a new process for deriving variations in the average surface magnetic field strength ($\B$) throughout time series of spectra. We validate our method and assess its performance on simulated spectra before applying it to archival observed spectra.
    Section~\ref{sec:observations} describes the observational data used in this study. Our method and its implementation are described in Sects.~\ref{sec:method} and~\ref{sec:implementation}. The application of our method to observed spectra is described in Sect.~\ref{sec:application-obs} and our results are discussed in Sect.~\ref{sec:conclusions}.

\section{Observations and data reduction}
\label{sec:observations}
This work takes advantage of public high-resolution spectroscopic observations obtained with three instruments: ESPaDOnS~\citep{donati-2006} and SPIRou~\citep{donati-2020}, installed at the Canada-France Hawaii-Telescope (CFHT), and Narval installed on the T\'elescope Bernard Lyot (TBL) at the Pic du Midi observatory, France~\citep{donati-2003}.

ESPaDOnS and Narval data were reduced with the \texttt{LIBRE-ESPRIT} reduction software~\citep{donati-1997b} and the continuum-normalized spectra were retrieved from \texttt{Polarbase}~\citep{petit-2014}.
SPIRou data were reduced with \texttt{APERO}~\citep{cook-2022} and downloaded from the Canadian Astrophysics Data Center (CADC). \texttt{APERO} provides multiple data products including spectra before and after correction of telluric lines (with extention \texttt{e.fits} and \texttt{t.fits}, respectively), along with blaze profiles estimated from flat-field exposures. An estimate of the photon noise was obtained from the number counts prior to telluric correction and used to estimate uncertainties in the telluric-corrected spectra. 

For this work, we refined our initial spectrum normalization procedure of SPIRou spectra, used by~\citet{cristofari-2023b, cristofari-2025b}. For each observation, the spectral orders were corrected by the blaze profiles and a 1D spectrum was obtained by merging all orders. In cases where spectral orders overlapped, only the segments with the highest throughput were kept to avoid biases of the normalization by noisy order edges. The blaze-corrected spectrum was then divided in windows of equal velocity width (which we empirically fixed to 3000\,$\kms$). In each window, we computed the flux 95th percentile and a cubic interpolation between the obtained points provided us with the continuum of the whole spectrum. Our process successfully captured low-frequency variation in the spectra, while preserving the structure of the spectral lines. We find that this process is particularly useful in preserving the shape of spectral lines recorded on an order edge, which can bias normalization in order-per-order normalization schemes.

Here, we consider three M~dwarfs: EV~Lac, DS~Leo, and Barnard's star. All their available SPIRou spectra were retrieved and combined for each observation night to achieve higher signal-to-noise ratios (S/Ns). Most of the observations recorded with SPIRou were obtained in the context of the SPIRou Legacy Survey. For EV~Lac and DS~Leo, we additionally retrieved all of the available Narval and ESPaDOnS data. Regions with deep telluric absorption bands were excluded from the analysis. Because both instruments have similar characteristics, observations were used together, and spectra were combined each night to obtain higher S/Ns. For EV~Lac, this yielded spectra for 67 nights between 2005 to 2016 and for DS~Leo, this yielded spectra for 61 nights between 2006 and 2014.

\section{Model description}
\label{sec:method}

The idea of extracting information about the stellar surface from relative variations in spectral features has been around for decades, with~\citet{gray-1991, gray-1994} selecting highly $\teff$-sensitive line ratios to derive temperature variations. Some recent studies have been focused on deriving relative radial velocity variations to achieve the meter-per-second precisions needed to detect and characterize exoplanets~\citep[see, e.g.,][]{artigau-2022, shahaf-2023}. More recently,~\citet{artigau-2024} proposed deriving relative temperature variations from a fully data-driven approach and relying on the formalism adopted by~\citet{bouchy-2001} and~\citet{artigau-2022}. We begin by briefly reviewing the basis of this formalism.

We consider a spectrum, $\vec{A}$, function of a parameter, $x$, which is assumed to have a single value for all pixels (e.g., effective temperature, metallicity, magnetic field intensity). Small variations between $\vec{A}$ and a reference spectrum, $\vec{A}_{\rm ref}$, can be approximated by the derivative of $\vec{A}$ with respect to $x$ evaluated at $x_{\rm ref}$, so that

\begin{equation}
        \frac{\vec{A}-\vec{A}_{\rm ref}}{x-x_{\rm ref}} = \left.\frac{\partial{\vec{A}}}{\partial x}(x)\,\right|_{x=x_{\rm ref}} 
        \label{eq:eq1}
\end{equation}

This approach provides a robust method to derive the variation of any quantity responsible for fluctuations in a spectrum as long as the derivative for each pixel is known and that the variations are small. \citet{artigau-2024} relied on this formalism to search for temperature variations, deriving $\partial{\vec{A}/\partial T}$ from grids of observed spectra.

In the context of magnetic fields, however, the implementation of this method leads to spurious results, which we attribute to two main factors. First, for most pixels, variations in flux as a function of magnetic field in the 0--10\,kG range are poorly modeled by a low-order polynomial; as a result, the assumption of small variations fails. Secondly, observed spectra are poorly modeled by a single magnetic field intensity~\citep[see, e.g.,][]{kochukhov-2021, reiners-2022,cristofari-2023, cristofari-2023b}.  
To overcome these limitations, it is necessary to adopt a more complete formalism describing the variation of pixels with magnetic field intensity for different magnetic components.

\subsection{Formalism}

In this section, we describe the now commonly used formalism~\citep[see, e.g.][]{shulyak-2014, shulyak-2019, reiners-2022, cristofari-2023, cristofari-2023b} that we adopted. Here, spectrum $\vec{A}$ is represented as
\begin{equation}
    \vec{A} = \sum{a_k\vec{F}_k}
,\end{equation}
where $\vec{F}_k$ is the synthetic disk-integrated spectrum of a star with an assumed radial magnetic field strength 
and $a_k$ is the filling factor associated with that component. Previous studies suggest a range of useful components for M dwarfs spanning $0$\,kG to $12$\,kG~\citep{shulyak-2019, reiners-2022, cristofari-2023}. Because the filling factors must add up to 1 for the problem to remain physical, this relation can be written as
\begin{equation}
    \vec{A} = \vec{F}_0+\sum{a_k(\vec{F}_k-\vec{F}_0)}
,\end{equation}
where $\vec{F}_0$ is the synthetic disk-integrated spectrum computed for a 0\,kG magnetic field. The synthetic spectrum $\vec{A}$ therefore depends solely on the filling factors, $a_k$, associated with each magnetic component.
Let us consider a reference spectrum, $\vec{A}_{\rm ref}$, described by the filling factors, $\epsilon_k$, so that ${\vec{A}_{\rm ref}=\vec{F}_0+\sum{\epsilon_k(\vec{F}_k-\vec{F}_0)}}$. Then a variation between $\vec{A}$ and $\vec{A}_{\rm ref}$ can be written as\begin{equation}
   \vec{A}-\vec{A}_{\rm ref} = \sum{(a_k-\epsilon_k)(\vec{F}_k-\vec{F}_0)}
    \label{eq:eq_linear_ff}
.\end{equation}
The variation in flux can be described by a variation in filling factors $a_k-\epsilon_k$, which we  denote as $\delta a_k$ hereafter. The variation in surface averaged magnetic field strength, $\B$, can then simply be obtained as $\delta\B = \sum\delta a_k B_k$, where $B_k$ is the magnetic field strength for the $k$-th component.

Solving  Eq.~\ref{eq:eq_linear_ff} allows us to derive $\delta a_k$, provided that the values of $A$, $A_{\rm ref}$, and $F_k$ for each magnetic components are known. In our application to observed spectra, the reference spectrum, $A_{\rm ref}$, was computed from the observed spectra directly, so the left hand-side of Eq.~\ref{eq:eq_linear_ff} comes from observations, while the right-hand side of the equation is informed by models (see Sect.~\ref{sec:implementation}).

\section{Implementation and simulations}
\label{sec:implementation}

Our implementation relies on synthetic spectra computed with \texttt{ZeeTurbo}~\citep{cristofari-2023}, which allowed us to obtain $\vec{F}_k-\vec{F}_0$ for different field strengths. Synthetic spectra were computed with higher sampling than observations, so that for an observed spectrum $\vec{A}$, we can choose the pixels in $\vec{F}_k-\vec{F}_0$ corresponding to the nearest observed wavelengths.
 We built the reference spectrum $\vec{A}_{\rm ref}$ by taking the median of the observed spectra in the stellar reference frame, which we refer to as the template hereafter. 
Our process is inherently sensitive to spurious pixels in the observations and flux fluctuations not properly accounted for by error bars (e.g., telluric line residuals). 
To mitigate the impact of spurious pixels, we quadratically added the standard deviation of the flux across the series of observed spectra to the photon noise associated with each pixel.
We then implemented an iterative process to reject bad pixels. Specifically, the weighted least-squares problem (Eq.~\ref{eq:eq_linear_ff}) can be solved using all points to obtain a first estimate of the $\delta a_k$, which is used to make a prediction on the variation of each pixel's flux. We then computed the reduced $\chi^2$ on the predicted and observed flux variations and used this value to inflate the error bars associated with each pixel. We rejected each pixel for which the flux ends up deviating by more than three times the inflated error bar and the weighted least-square problem was solved on the remaining pixels. The process was repeated until all spurious pixels are rejected. Convergence was typically reached within the first ten iterations in our applications. This approach provides a robust and conservative way to reject bad pixels that could impact the results. We find that this method is particularly robust to provide estimates for spectra with significant artifacts.

\subsection{Application to simulated spectra}
We constructed a series of "synthetic observations," relying on synthetic spectra computed with \texttt{ZeeTurbo}. We relied on the results from~\citet{cristofari-2025b} to extract realistic filling-factor distributions for the relevant series of observation nights. Consequently, we built synthetic observations from linear combinations of models computed for magnetic field strengths ranging from 0 to 10\,kG in steps of 2\,kG. Additionally, we relied on temperature variation estimates ($\delta T$) obtained with the LBL package~\citep{artigau-2024} from the same data to simulate the impact of temperature variations at the stellar surface, which have been shown to be significantly anti-correlated with surface averaged magnetic field variations ($\delta \B$) for the stars considered in this work~\citep{cristofari-2025b}. The modeled spectra were convolved with a Gaussian profile of full width at half maximum (FWHM) of 4.3\,$\kms$ (to mimic the instrumental width of SPIRou), followed by a second Gaussian profile of 5.0\,$\kms$ to represent the typical broadening associated with rotation and macroturbulence for a star such as EV~Lac (see Sect.~\ref{sec:application-obs}). We note that convolutions can only approximately reproduce the impact of rotation or macro-turbulence on spectra, especially for lines with complex Zeeman patterns. The current method, however, relies on the relative variation of fluxes as a function of field strength, mitigating the impact of the approximation. Relying on convolution enables faster computations and our method proves relatively insensitive to broadening (see Sect.~\ref{sec:implementation-atm}). We note that detailed disk integration could be added in future implementations. The broadened synthetic spectra were then re-binned on a wavelength grid typical for spectra recorded with SPIRou or Narval.
Additional series of models were obtained by considering either magnetic field variations or temperature variations alone. 
Our process was applied to the synthetic observations, relying on different assumptions to assess the behavior of our method and its sensitivity to various parameters.

\subsection{Sensitivity to normalization}
\label{sec:normalization}

\begin{figure}
        \centering
        \includegraphics[width=\linewidth]{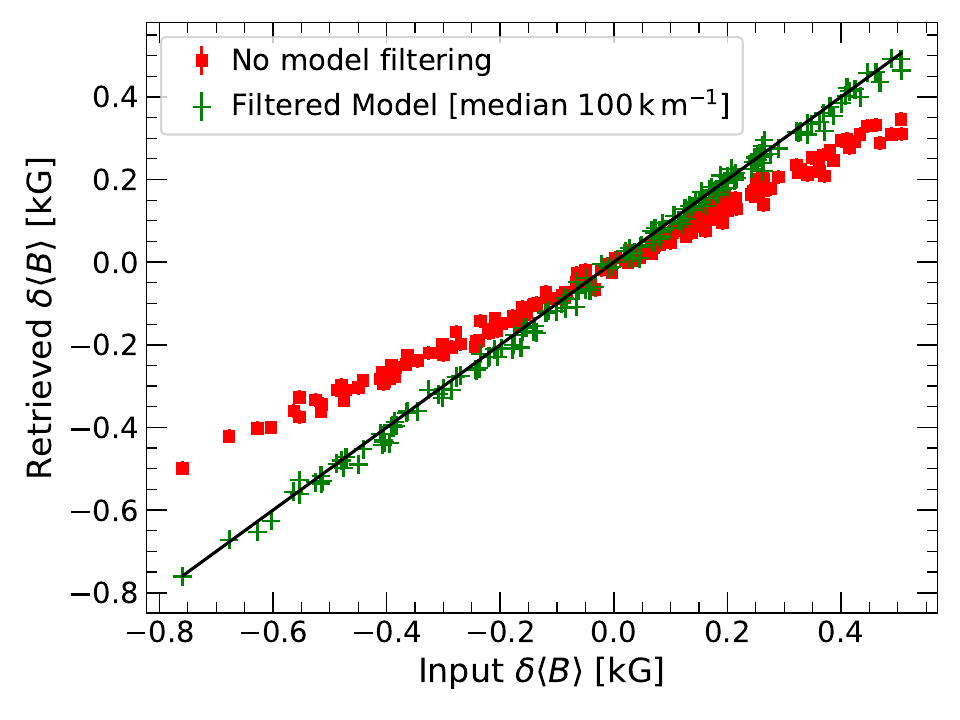}
        \caption{Comparison between the retrieved (y-axis) and input (x-axis) relative magnetic field strengths ($\delta\B$). 
                 The synthetic observations were normalized to the template using a rolling median with a window of 100\,$\kms$.
                 The red squares and green plus symbols (`+') show the results obtained before and after re-normalizing the models used to computed $\vec{F}_k-\vec{F}_0$ (Eq.~\ref{eq:eq_linear_ff}) with a 100\,$\kms$ median filter, respectively (see text).
                 The black line marks the equality.
                 This figure illustrates the need for careful re-normalization of the models to avoid biases in the results. 
                 }
        \label{fig:damping-norm}
\end{figure}

Low-frequency fluctuations are often observed in echelle spectra, with the potential to lead to biased $\delta\B$ and $\delta T$ estimates. While an absolute normalization of the continuum is not necessary, relative normalization with respect to the reference spectrum is crucial to obtain accurate results with the proposed method. 

To mitigate the impact of low-frequency continuum fluctuations, a moving median with a width of 500\,$\kms$ was applied to the residuals between each observed spectrum and the template. The resulting continuum variations were used to correct each individual observation. We tested the impact of such a normalization step by applying it to our series of synthetic observations. 
Specifically, we applied rolling medians with different window sizes to synthetic observations and applied our method with and without filtering the models used to compute $\vec{F}_k-\vec{F}_0$.
Our simulations reveal that if our process is applied to median-filtered observations without filtering the models, the amplitude of the retrieved fluctuations are "damped", meaning that the structure of the variations is conserved, but their amplitude is reduced (see Fig.~\ref{fig:damping-norm}, red squares). This effect can be mitigated by filtering the models used to compute $\vec{F}_k-\vec{F}_0$ with a rolling median of similar window size (see Fig.~\ref{fig:damping-norm}, green + symbols). We note that such filtering is not perfect, in particular when applied to spectral regions containing broad or blended lines.
We found that applying a rolling median to the residuals between the models and model of reference (e.g., to $\vec{F}_k-\vec{F}_0$ directly) provides the most robust results, as this limits the changes in pseudo-continuum caused by broad absorption features. 
Furthermore, we note that strong magnetic fields can significantly broaden the absorption features, fundamentally changing the shape of observed pseudo-continuum. Consequently, one may favor larger windows for the rolling median to avoid introducing spurious feature variations in the models.

\subsection{Relation between temperature and magnetic fields}
\label{sec:implementation-teff}

Recent results have provided strong evidence of anti-correlations between the obtained small-scale field measurements and temperature variations ($\delta T$) for a few strongly magnetic targets~\citep{artigau-2024, cristofari-2025b}. In this work, we investigate the impact of temperature variations on magnetic field estimates and vice versa.

To retrieve the temperature variations, we implemented a process similar to that described by~\citet{artigau-2024}, with the notable difference that we derived $\partial \vec{A}/\partial T$ for each pixel from our synthetic spectra. To validate our process, we created a series of modeled observations with varying effective temperatures, following the estimate obtained with the LBL approach for EV~Lac (with typical variations of less than 20\,K). Our implementation allows us to successfully retrieve the temperature variations used to generate the models. 

\begin{figure}
        \includegraphics[width=\linewidth]{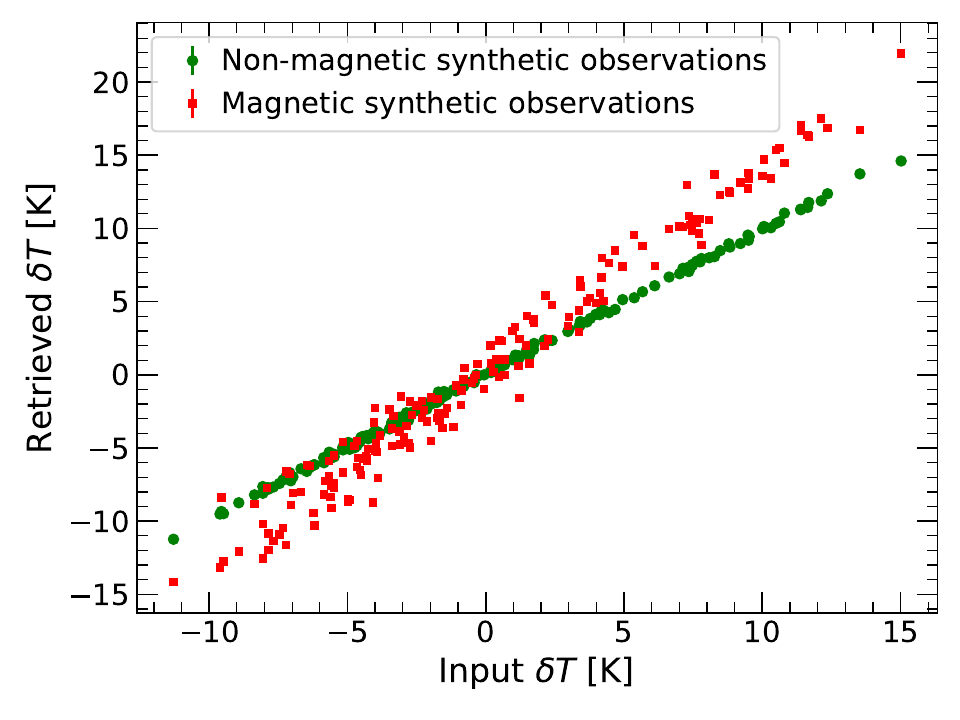}
        \caption{Comparison between the retrieved $\delta T$ (y-axis) and the input $\delta T$ used to generate the synthetic observations (x-axis). Synthetic observations were computed on the wavelength range covered by Narval/ESPaDOnS, accounting for magnetic field variations (red squares), or fixing the value of the magnetic fields to 0\,kG (green circles). This figure illustrates how magnetic fields can impact the $\delta T$ estimates.}
        \label{fig:impact-b-on-teff}
\end{figure}

To investigate the impact of magnetic fields on temperature variation estimates, we modeled the spectra, accounting for both temperature and magnetic fields variations by taking realistic estimates for EV~Lac from~\citet{cristofari-2025b}. The average surface magnetic field strengths then ranged from 4 to 5\,kG, and the effective temperatures range from 3290 to 3310\,K.
We found that temperature variations have a limited impact on the derived $\delta\B$. The inclusion of magnetic fields, however, was shown to have a much more significant impact on the derived $\delta T$ (see Fig.~\ref{fig:impact-b-on-teff}). We found that this impact is particularly significant for simulations covering the Narval/ESPaDOnS wavelength domain, where a very large number of atomic features are predicted to be sensitive to magnetic fields.

The results of this experiment show that $\delta T$ and $\delta \B$  estimates are not completely independent, with a greater sensitivity of $\delta T$ estimates to magnetic field variations seen for realistic conditions. We note that typical $\delta T$ are small, with a few kelvin variations leading to small changes in the observed spectra, while $\delta\B$ fluctuations in stars like EV~Lac can lead to more visible pixel variations. Furthermore, magnetic field variations in stellar spectra lead to effects that are strongly dependent on the considered spectral line. Having both magnetically sensitive and insensitive lines in the spectral domain makes it possible to obtain robust constraints on $\delta\B$.
A comparison between the model-driven and data-driven $\delta T$ estimates is discussed in Sect.~\ref{sec:application-dtemp}. 

\subsection{Impact of atmospheric parameters on $\langle B \rangle$ variations}
\label{sec:implementation-atm}

We investigated the impact that incorrect assumptions on the atmospheric properties of the star can have on the results. To achieve this, modeled observations were analyzed with $\vec{F}_k-\vec{F}_0$ computed for different effective temperatures ($\teff$), surface gravities ($\log{g}$), metallicities ($\rm [M/H]$), and projected rotational velocity ($v\sin{i}$).

We find that an offset in atmospheric parameters has a `damping' effect on the retrieved relative field strengths, similar to the impact of short-window normalization (see Sec~\ref{sec:normalization}): the structure of the variations are typically conserved, while the amplitude of the field variations are either lower or higher than expected. Through simulations, we find that a change of $300$\,K can yield a maximum difference in $\delta\B$ of 0.1\,kG in the case of EV~Lac (whose magnetic field is able to change by up to 1\,kG in our simulations), comparable to the typical uncertainties reported in~\citet{cristofari-2025b}. 
Similarly, changes of 0.5\,dex in the assumed surface gravity or metallicity can lead to maximum differences in $\delta \B$ of $0.2$\,kG. 
Changes in the assumed broadening of the spectra of up to 5\,$\kms$ typically lead to maximum changes in $\delta \B$ of $0.05$\,kG. 
The typical uncertainties on $\teff$, $\log{g}$, $\rm [M/H]$ and $v\sin{i}$ are generally lower than 300\,K, 0.5\,dex or 5\,$\kms$, while the maximum differences in $\delta\B$ are then lower than the typical uncertainties associated with the measurements.
Our process remains therefore robust to changes in atmospheric parameters and broadening, which allows us to limit the impact of incorrect stellar characterization on the results.
 
\section{Application to observations}
\label{sec:application-obs}
\subsection{Application to SPIRou spectra}

We applied our process to spectra recorded with SPIRou for EV~Lac, DS~Leo, and Barnard's star, whose small-scale magnetic fields were previously characterized from fits of synthetic spectra to each individual observation~\citep{cristofari-2025b}. This allowed us to make direct comparisons between the obtained $\delta\B$ and the magnetic field estimates reported by~\citet{cristofari-2025b}.
For each star, we assumed atmospheric parameters that were close to those obtained by~\citet{cristofari-2025b}, which we list in Table~\ref{tab:ref-parameters}. The broadening parameters (i.e., turbulence, rotation, instrumental width) were approximated by a single Gaussian profile. While the finetuning of each of these parameters is possible, we chose to keep the assumed atmospheric parameters and broadening kernels approximate, given the relative insensitivity of the proposed method to detailed stellar characterization (see Sect.~\ref{sec:implementation-atm}). 

\begin{table}
\caption{Atmospheric parameters adopted in the current work.}
\label{tab:ref-parameters}
        \begin{tabular}{ccccccccc}
                \hline
                \hline
        Star & Atmospheric & Adopted & \\
        & parameters & parameters \\
        \hline
        EV~Lac & $\teff=3342$\,K & $\teff=3300$\,K \\
        (Gl~873) & $\log{g}=4.75$\,dex & $\log{g}=5.0$\,dex \\
        & ${\rm [M/H]}=0.02\pm0.10$\,dex & ${\rm [M/H]}=0.0$\,dex \\
        & ${v_{\rm eq}\sin{i}}=3.0$\,$\kms$ & \multirow{2}*{$v_{\rm b}=5.0\,\kms$}  \\
        & ${\zeta_{\rm RT}}=4.22$\,$\kms$ &\\
        \hline
        DS~Leo & $\teff=3797\pm30$\,K & $\teff=3800$\,K \\
        (Gl~410) & $\log{g}=4.67\pm0.05$\,dex & $\log{g}=4.5$\,dex \\
        & ${\rm [M/H]}=-0.02\pm0.10$\,dex & ${\rm [M/H]}=0.0$\,dex \\
        & ${v_{\rm eq}\sin{i}}=1.5$\,$\kms$ & \multirow{2}*{$v_{\rm b}=3.0\,\kms$}  \\
        & ${\zeta_{\rm RT}}=3.01$\,$\kms$ &\\
        \hline
        Barnard & $\teff=3300\pm31$\,K & $\teff=3300$\,K \\
        (Gl~699) & $\log{g}=4.71\pm0.06$\,dex & $\log{g}=5.0$\,dex \\
        & ${\rm [M/H]}=-0.54\pm0.10$\,dex & ${\rm [M/H]}=-0.5$\,dex \\
        & ${v_{\rm eq}\sin{i}}=<0.1$\,$\kms$ & \multirow{2}*{$v_{\rm b}=4.0\,\kms$}  \\
        & ${\zeta_{\rm RT}}=3.79$\,$\kms$ &\\
        \hline
    \end{tabular}
 \tablefoot{The second column shows the parameters reported by~\citet{cristofari-2025b}. The third column presents the approximate parameters adopted for our work, corresponding the closest point in our grid of models.}
\end{table}

\begin{figure}
    \centering
    \includegraphics[width=0.99\linewidth]{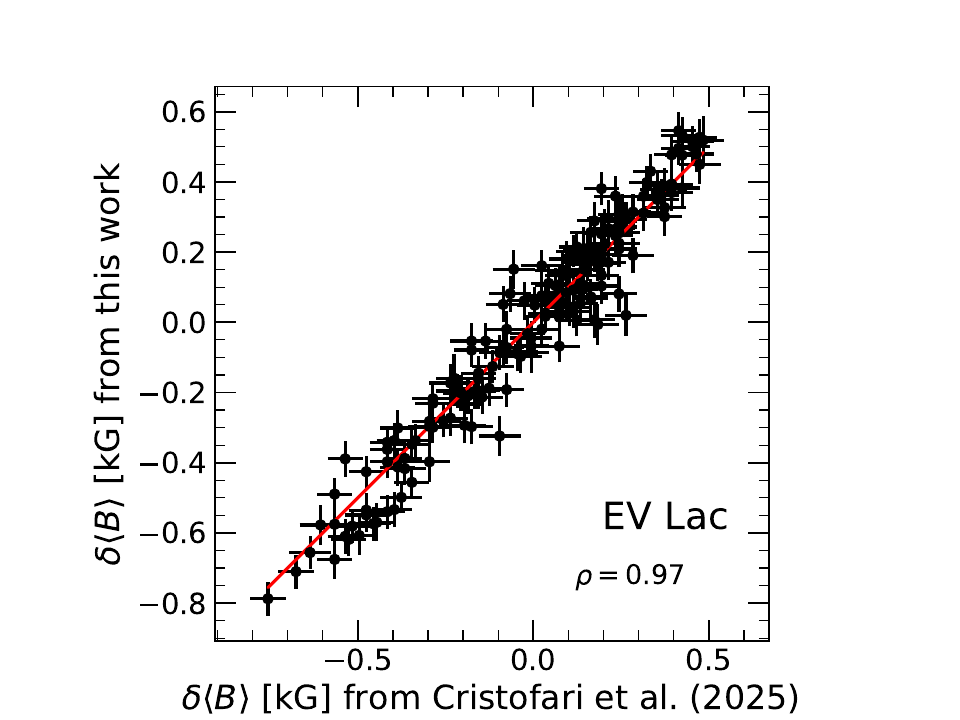}
    \caption{Comparison between $\delta\B$ obtained in this work and the values of~\citet{cristofari-2025b}. The average $\B$ was removed from the estimates of~\citet{cristofari-2025b} for comparison. The red line shows the equality. The Pearson correlation coefficient ($\rho$) is shown on the figure. Our process allows us to retrieve $\delta\B$ values that are in very good agreement with those obtained from fits of synthetic spectra to the data.}
    \label{fig:compare-gl873}
\end{figure}

For all three stars, we obtained magnetic field variations that were in very good agreement with the results of~\citet[as also seen in our Figs.~\ref{fig:compare-gl873} and~\ref{fig:compare-gl410+gl699}]{cristofari-2025b}, with reduced $\chi^2$ values between the two series of points of 1.01, 1.07, and 1.26. In addition, we obtained Pearson correlation coefficients of $0.97$, $0.89,$ and $0.58$ for EV~Lac, DS~Leo, and Barnard's star, respectively.
Relying on the same tools as~\citet{cristofari-2025b}, we fit a quasi-periodic Gaussian process (QPGP) to the retrieved $\delta\B$ for each star. 
Specifically, we relied on a GP regression with a kernel, previously used in other studies~\citep{angus-2018, fouque-2023, cristofari-2025b}, expressed as
\begin{equation}
        \kappa(t_i, t_j)=\alpha^2\exp\Big[-\frac{(t_i - t_j)^2}{2l^2}-\frac{1}{2\beta^2}\sin^2\Big(\frac{\pi(t_i - t_j)}{P_{\rm rot}}\Big)\Big] + \sigma^2\delta_{ij}
        \label{eq:equation}
,\end{equation}

\noindent where $\alpha$ is the amplitude of the GP, $\beta$ a smoothing factor, $l$ the decay time,  $t_i-t_j$  the time difference between two data points, i and j, and $\sigma$ the standard deviation of an added uncorrelated white noise. Finally, $P_{\rm rot}$ is the recurrence period, which we identify as the rotation period of the star.

The posterior distributions provide constraints on the rotation period of the star that are in excellent agreement with the results of~\citet{cristofari-2025b} and similar long-term fluctuations were also observed.
These results demonstrate our ability to estimate $\B$ variations that are consistent with those obtained from fit of synthetic spectra to each individual observation (see Table~\ref{tab:gp-spirou} and Figs.~\ref{fig:spirou-gl873-corner},~\ref{fig:spirou-gl410-corner}, and~\ref{fig:spirou-gl699-corner}). Since the proposed method to derive $\delta\B$ relies on solving the linear Eq.~\ref{eq:eq_linear_ff} and does not require us to explore the parameter space with a Markov chain Monte Carlo (MCMC), these results can be obtained  faster by several orders of magnitude than the results derived by~\citet{cristofari-2025b}.

\subsection{Application to ESPaDOnS and Narval spectra}

\begin{figure}
    \centering
    \includegraphics[width=1.0\linewidth]{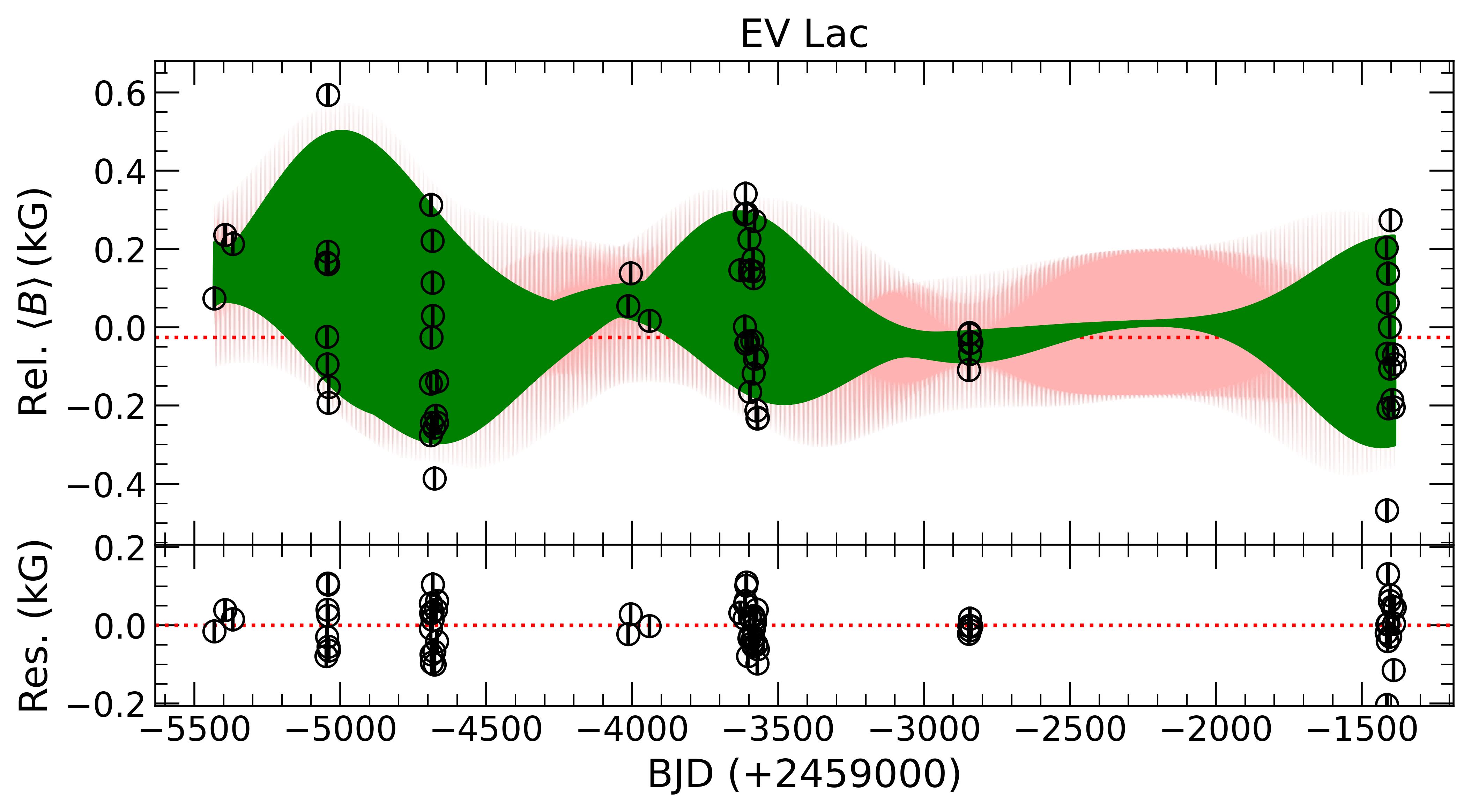}
    \includegraphics[width=1.0\linewidth]{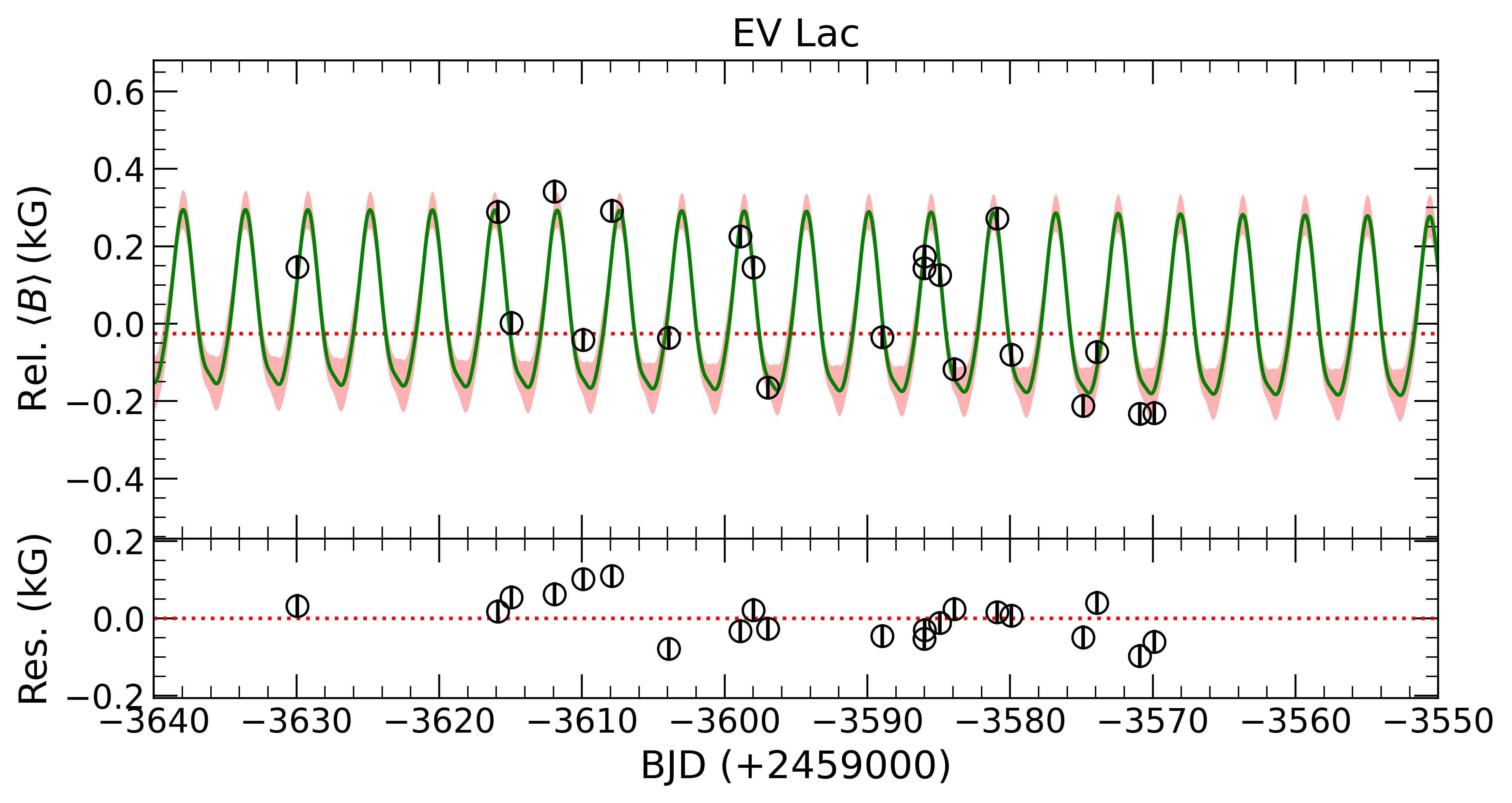}
    \includegraphics[width=1.0\linewidth]{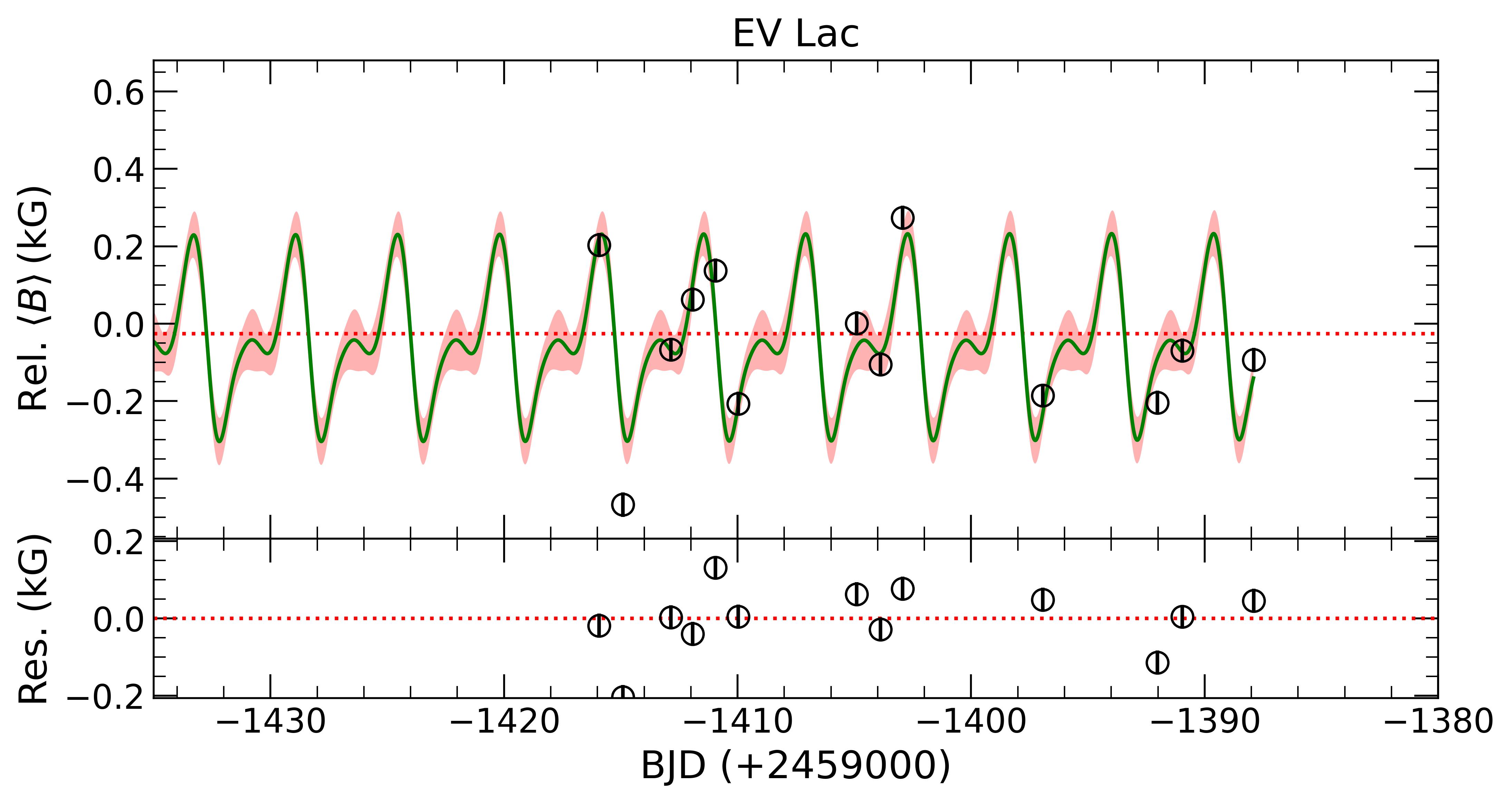}
    \caption{Best QPGP fit (green) obtained on relative small-scale magnetic field estimates derived from ESPaDOnS and Narval spectra for EV~Lac. Top panel: Fit over the entire dataset. Middle and bottom: Segments of the dataset. The residuals (Res.) between data points and GP fit are shown on each panel.}
    \label{fig:narval-gl873-gp}
\end{figure}

\begin{figure}
        \centering
        \includegraphics[width=1.0\linewidth]{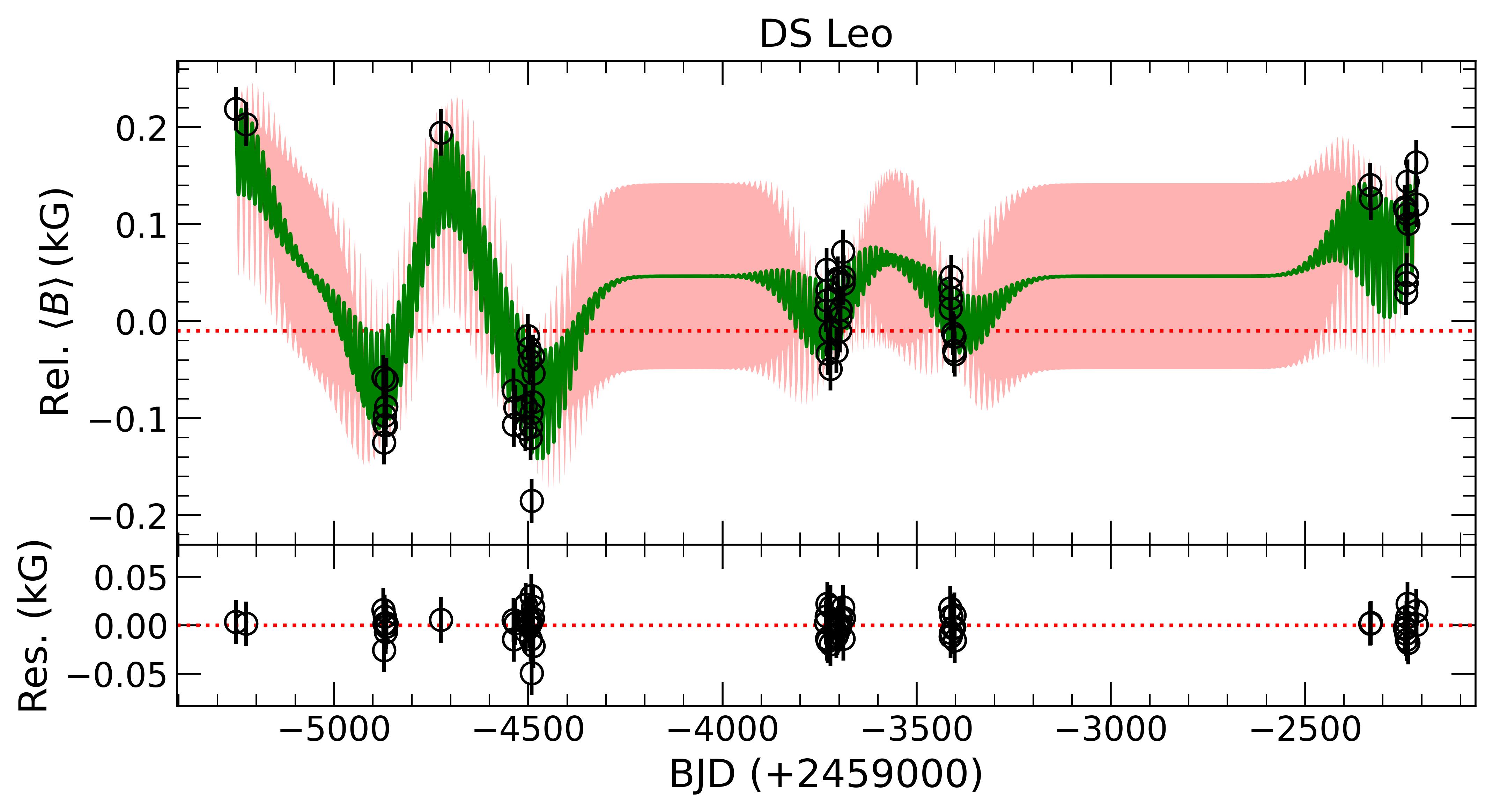}
        \includegraphics[width=1.0\linewidth]{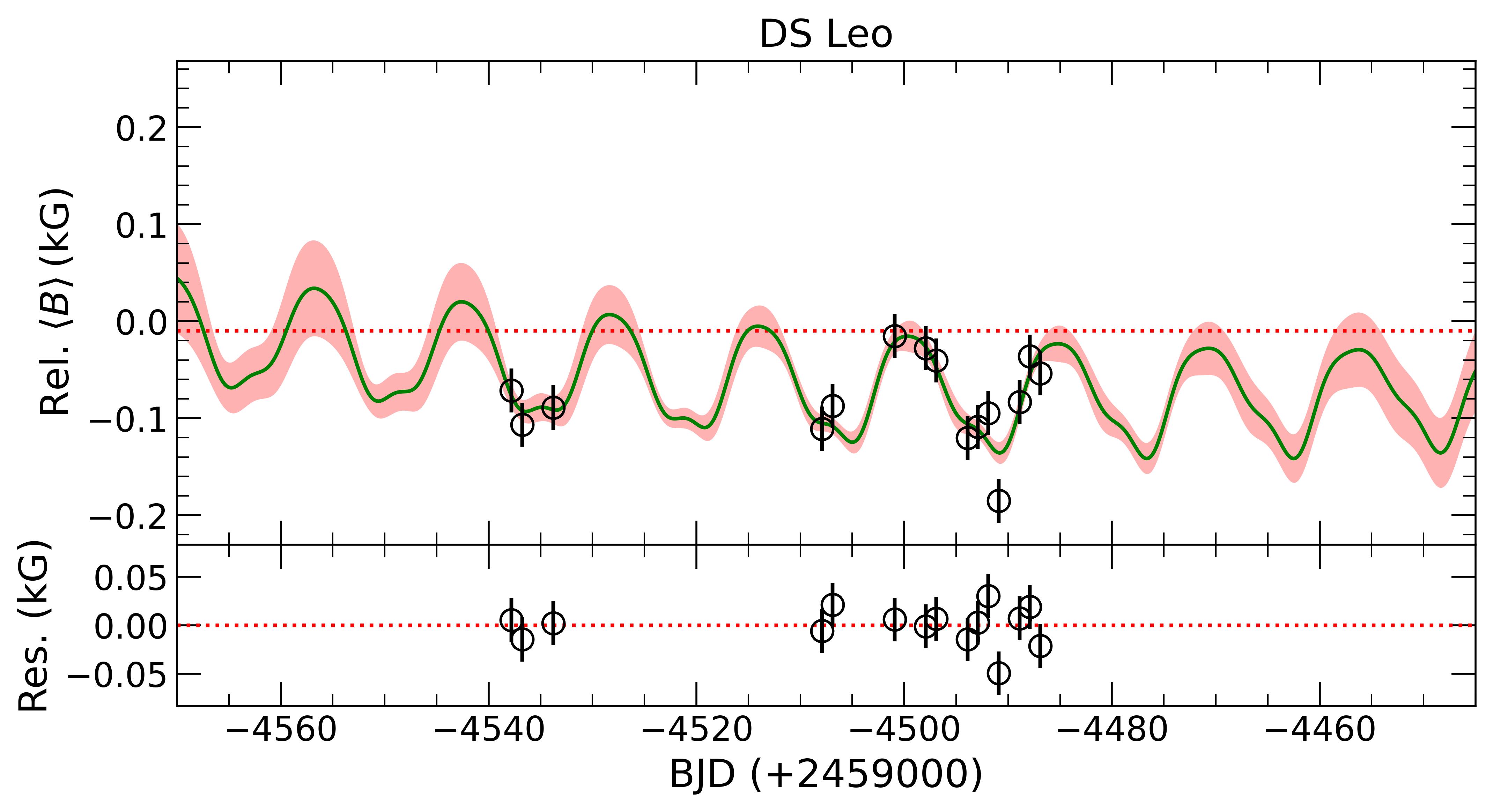}
        \includegraphics[width=1.0\linewidth]{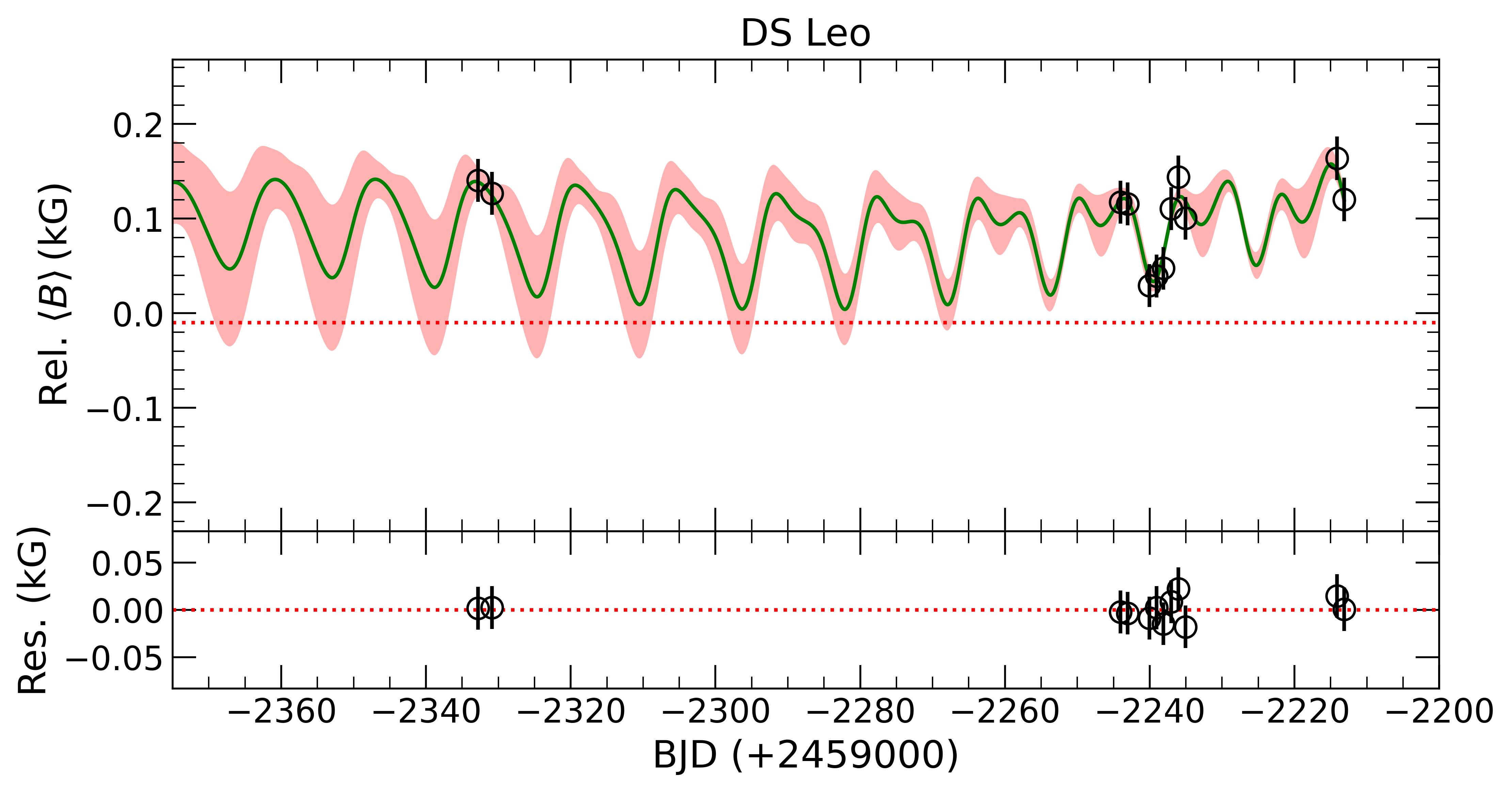}
        \caption{Same as Fig.~\ref{fig:narval-gl873-gp}, but for DS~Leo.}
        \label{fig:narval-gl410-gp}
\end{figure}

To complete our analysis, we applied our process to optical spectra recorded with Narval and ESPaDOnS for EV~Lac and DS~Leo (see Sect.~\ref{sec:observations}). The derived relative magnetic field strengths are well modeled by a QPGP (see Figs.~\ref{fig:narval-gl873-gp} and~\ref{fig:narval-gl410-gp}), yielding rotation periods of $P_{\rm rot}= 4.367\pm0.023$\,d and $P_{\rm rot}= 14.105\pm0.406$\,d for EV~Lac and DS~Leo, respectively. These estimates are in very good agreement with those reported by~\citet{cristofari-2025b} of  $P_{\rm rot}= 4.362\pm0.001$\,d and $P_{\rm rot}= 13.953 \pm0.088$\,d, albeit with larger error bars. The larger error bars can partly be attributed to the sparsity of the available  Narval and ESPaDOnS data, as a result of the observing strategies primarily focused on Zeeman Doppler Imaging reconstructions~\citep[see, e.g.,][]{morin-2008, morin-2011, donati-2008, bellotti-2024}. This underscores the benefits of long-term spectroscopic monitoring of M~dwarfs, not only for the planet search, but for the study of their magnetic fields as well.
We note that DS~Leo was reported to exhibit differential rotation, with distinct rotation periods at the equator and the poles~\citep[$13.37\pm0.86$\,d and $14.96\pm1.25$\,d, respectively;][]{hebrard-2016}, which could in part account for the larger error bar derived with our GP fit.

\subsection{Model-driven $\delta T$ variations}
\label{sec:application-dtemp}

We applied the model-driven approach used to derive temperature variation (see Sect.~\ref{sec:implementation-teff}) to the observed spectra considered in our study.
For EV~Lac (Gl~873) and DS~Leo (Gl~410), we find significant correlations between our derived $\delta T$ and the data-driven estimates obtained with the LBL~\citep{artigau-2024} and published by~\citet{cristofari-2025b}, with Pearson correlation coefficients of 0.95 and 0.93, respectively (see Fig.~\ref{fig:compare-dtemp-gl873+gl410+gl699}).
A moderate correlation is also observed for Barnard's star, with a Pearson coefficient of 0.59 (see Fig.~\ref{fig:compare-dtemp-gl873+gl410+gl699}). We note the smaller range of the variations relative to error bars for this star.
These results demonstrate the ability models have to reproduce relative temperature variations in spite of the significant discrepancies between models and observations (see, e.g.,~\citealt{cristofari-2022, cristofari-2022b}). We note that in a similar experiment,~\citet{artigau-2024} found that model-driven variations led to a damping of the retrieved $\delta T$. Details in implementation and normalization, along with differences between the models used could explain our ability to retrieve amplitudes comparable to those obtained with the data-driven approach. Our results further support previous findings~\citep{bellotti-2025} where synthetic spectra were used to estimate temperature and $\B$ variations from spectra recorded with SPIRou.

\begin{figure}
        \includegraphics[width=\linewidth]{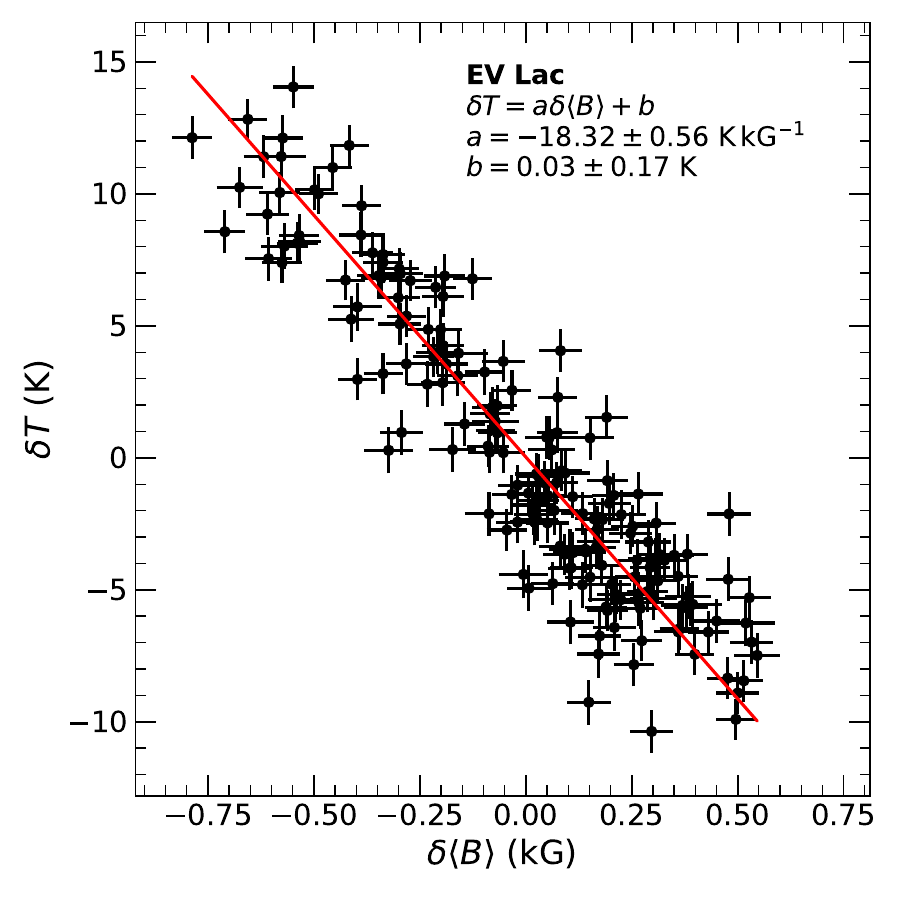}
        \caption{Relation between our model-driven $\delta T$ and $\delta \B$ estimates obtained from SPIRou 
                data for EV~Lac. The red line shows the best linear fit to the data point, obtained with an orthogonal distance regression. The parameters of the fit are shown on the figure.}
        \label{fig:correlation_teff_b_spirou_gl873}
\end{figure}

From the spectra recorded with SPIRou, we observed strong anti-correlations between our $\delta T$ and $\delta\B$ estimates. We applied orthogonal distance regressions\footnote{Implementation relying on Scipy~\citep{scipy}} (ODRs) to estimate the properties of the linear relationships between $\delta T$ and $\delta\B$ (see Fig.~\ref{fig:correlation_teff_b_spirou_gl873} and~\ref{fig:correlation_teff_b_spirou_gl410+gl699}). 
For EV~Lac, DS~Leo, and Barnard's star, the fits yield slopes of $-18.3\pm0.6$, $-54.2\pm1.8$\,$\rm K\,kG^{-1}$, and $-24.7\pm1.0$\,$\rm K\,kG^{-1}$, consistent with those reported by~\citet[][$-19.8\pm0.5$, $-60.4\pm2.5,$ and $-18.8\pm0.8$\,$\rm K\,kG^{-1}$, respectively]{cristofari-2025b}. 
We note that the ODR is very sensitive to the relative scaling of the error bars of both axes, which can have an impact of several $\rm K\,kG^{-1}$ in our case.

 From the data recorded with Narval/ESPaDOnS, we can observe similar anticorrelations between $\delta T$ and $\delta\B$, with significantly larger scatters, yielding slopes of  $-22.3\pm3.3$ and $-61.7\pm6.8$\,$\rm K\,kG^{-1}$ for EV~Lac and DS~Leo, respectively.
  The larger scatters in the results could be attributed to the impact of magnetic fields on $\delta T$ estimates (see Sec~\ref{sec:implementation-teff}), the sparsity of the observations obtained with Narval/ESPaDOnS, or the higher contrast in the optical where cool spots could lead to larger temperature variations which may not be strictly anticorrelated with $\delta B$.

We note that while our simulations suggest that the impact of magnetic fields on the derived $\delta T$ is more significant for the Narval/ESPaDOnS domain, it could still have a non-negligible impact on the estimates obtained from spectra recorded in the nIR, which could lead to steeper slopes and larger scattering.
A natural next step would be to use the $\delta\B$ estimates to obtain magnetic field-dependent $\partial \vec{A}/\partial T$ values (see Eq.~\ref{eq:eq_linear_ff}) and rely on these to refine the $\delta T$ measurements. We leave the implementation of this process to a future study.

\subsubsection{Comparison to activity indicators}

From Narval and ESPaDOnS spectra, we measured the S-index, commonly used as an activity proxy, to compare it to our $\delta\B$ estimates.
We followed the definition of the S-index from~\citet{vaughan-1978}:
\begin{equation}\text{S-index} = \frac{aF_H+bF_K}{cF_R+dF_V}+e\end{equation}
where $F_H$ and $F_K$ are the fluxes taken in $1.09\,\AA$ triangular bands centered on $3968.470$ and $3933.661\,\AA$, respectively. $F_R$ and $F_V$ are the fluxes computed in $20\,\AA$ rectangular bands centered on $3901$ and $4001\,\AA$, respectively (see Fig.~\ref{fig:example_bands_sindex}). The coefficients $a$, $b$, $c$, $d$, and $e$, which allow us to calibrate the index to the Mount Wilson scale, were taken from~\citet{marsden-2014}. The flux uncertainties were propagated to estimate the error bars on each band's flux, which were, in turn, used to define error bars on the S-index measurements.

\begin{figure}
        \includegraphics[width=\linewidth]{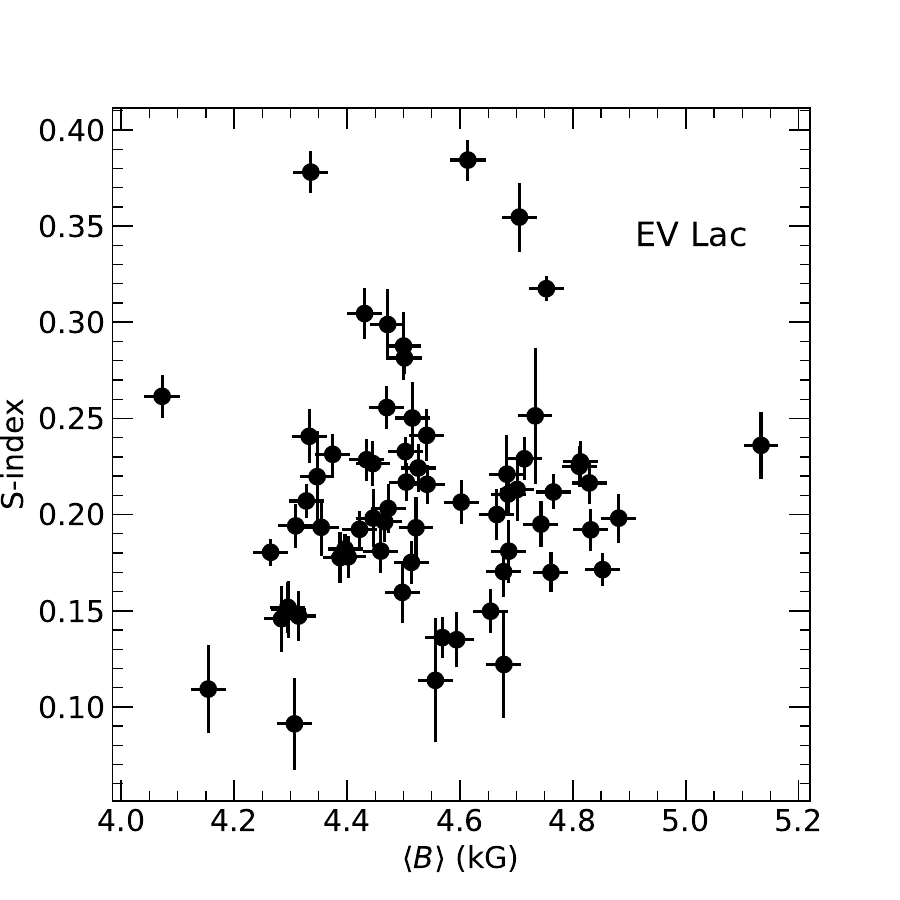}
                \includegraphics[width=\linewidth]{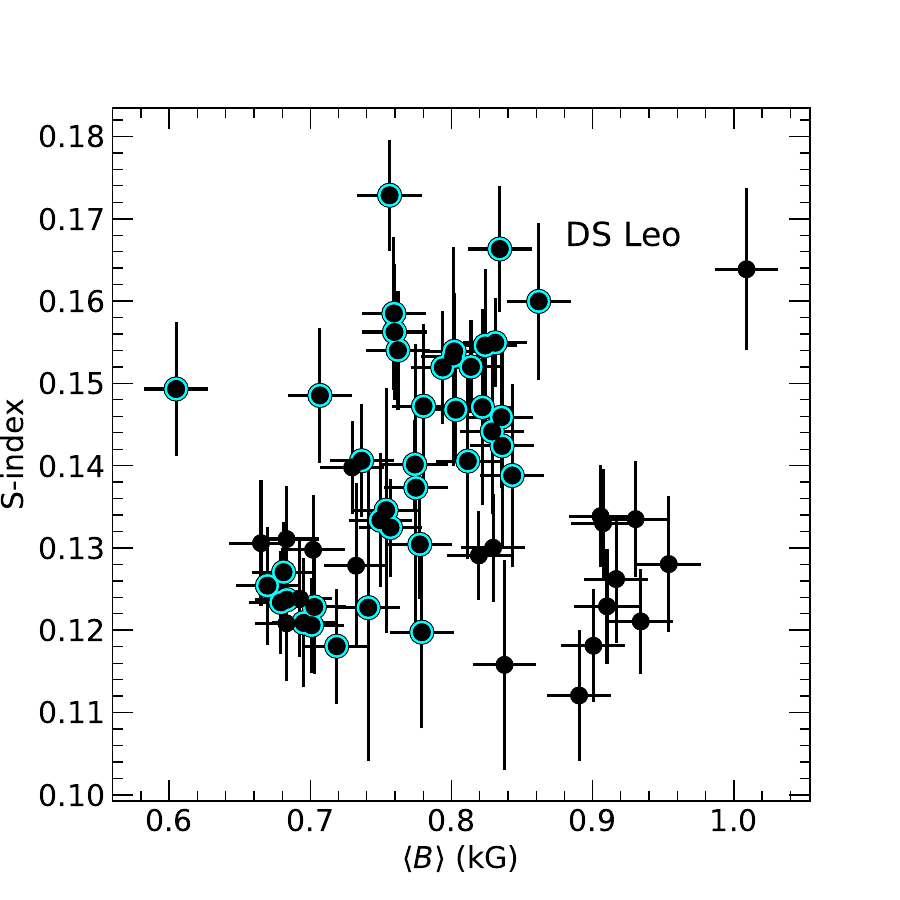}
        \caption{S-index compared to $\B$ for EV~Lac (top) and DS~Leo (bottom). $\B$ estimates were obtained from $\delta\B$ by adding typical median values from~\citet{cristofari-2025b}, of 4.54 and 0.79\,kG for EV~Lac and DS~Leo, respectively.  For DS~Leo, points marked with a cyan circle correspond to observations recorded between BJD 2454400 and 2455700 (see Fig.~\ref{fig:narval-gl410-gp}).}
        \label{fig:sindex-gl873}
\end{figure}

We recall here that $\delta\B$ is directly linked to $\B$ such that $\B=\delta\B+\B_{\rm ref}$, with $\B_{\rm ref}$ the value associated with the reference spectrum (see Eq.~\ref{eq:eq_linear_ff}). For an easier comparison with the literature, in Fig.~\ref{fig:sindex-gl873}, we computed $\B$ from $\delta\B,$ taking the typical $\B_{\rm ref}$ values from~\citet{cristofari-2025b} for EV~Lac and DS~Leo ($4.54$ and $0.79$\,kG, respectively).  No clear correlation was observed between the measured S-index and the magnetic field strength variation derived with our approach for EV~Lac or DS~Leo  (Fig.~\ref{fig:sindex-gl873}). We note that the long baseline of the observations ($\sim 10$ yrs) could encompass activity regime changes that could break the activity-$\B$ relationship that  would be expected for an activity driven by quiescent heating. 
For both stars, we divided the observations in segments of closely spaced observations and compared the activity indicators to $\delta\B$ in these. 
A tentative correlation (with a Pearson coefficient of 0.57) can be observed for DS~Leo if we are only considering data points between BJD 2454400 and 2455700. These dates encompass observation campaigns where the magnetic field strength is at its lowest. These results further suggest that changes in activity regimes could throw off the relation between the S-index and average surface magnetic field strength on long observation baselines.
Recently,~\citet{hahlin-2026} reported correlations between the S-index and $\B$ for a few hotter stars ($T_{\rm eff}$ ranging from $\sim$\,4350 to 5400\,K), suggesting that such a relation might be harder to recover for M dwarfs than it would be for Sun-like stars.

Moreover, DS~Leo and EV~Lac are known to exhibit flares~\citep[e.g.,][]{muheki-2020, duvvuri-2023} with a direct impact on activity indicators that can throw off the relation between magnetic fields and activity~\citep{getman-2025}. 
Previous work noted that flares can lead to larger X-ray luminosity than expected for a given magnetic flux~\citep{pevtsov-2003, getman-2025}, suggesting that much more energy is released by flares than by quiescent magnetic heating. These flares can impact all outer layers of the stellar atmosphere, consequently enhancing chromospheric emission.
These results underscore the importance of average magnetic field measurements, less prone to biases associated with chromospheric activity.

\section{Discussion and conclusions}
\label{sec:conclusions}
    In this paper, we study the temporal behavior of magnetic fields from long-term spectroscopic monitoring of cool stars~\citep[see, e.g.,][]{donati-2023, cristofari-2025b}. We explore a simple method to estimate relative variations of small-scale magnetic fields ($\delta\B$) from high-resolution spectra. The proposed method does not rely on fits of synthetic spectra to individual observations, providing extremely fast estimates in comparison.

    Studies that rely on fits of synthetic spectra to observations~\citep[e.g.,][]{reiners-2022, cristofari-2025b} are inherently limited by the quality of the fits, with systematics between modeled and observed line shapes leading to uncertainties and potential biases in the results. The method proposed in this paper attempts to mitigate these effects by only studying changes in line shapes relative to a reference spectrum, obtained from the observations themselves. While the fundamental idea of studying spectral feature variations compared to a reference spectrum is not new, with notable applications for radial velocity~\citep{artigau-2022} and surface temperature variations~\citep{artigau-2024}, this is to our knowledge the first such attempt to derive relative small-scale magnetic field variations ($\delta\B$).
    
    With this work, we show that our model-informed process applied to observations can provide reliable information on $\delta\B$ from series of high-resolution spectra.   
    Our approach relies on synthetic spectra computed with \texttt{ZeeTurbo}~\citep{cristofari-2023},  applied to high-resolution near-infrared and optical spectra recorded with SPIRou, Narval, and ESPaDOnS.
    Our results were compared to magnetic field measurements previously obtained for a few strongly magnetic targets~\citep{cristofari-2025b}, yielding an excellent agreement. 
    This underscores the reliability of the process, which (by construction) allows us to obtain $\delta\B$ from a linear fit, instead of expensive MCMC sampling~\citep{cristofari-2025b}. Consequently, the proposed method enables the treatment of hundreds of spectra in seconds, making it a diagnostic tool that is viable for systematic application to spectra recorded in the context of large surveys to come.

    The analysis of Narval and ESPaDOnS spectra provide new constraints on the long-term variations of the small-scale magnetic fields for EV~Lac and DS~Leo. QPGP fits to the series of $\delta\B$ measurements show a clear modulation consistent with the rotation period of the stars. We note that this has been achieved in spite of the relative sparsity of the data over a ten-year period, compared the dense survey carried out with SPIRou over three years.

    Through simulations and analysis of the observed data, we investigated the impact that assumptions on atmospheric properties have on the results and found that our method is relatively resilient with respect to small changes in the adopted parameters. 
    We also found that significant offset in the assumed atmospheric properties leads to a damping of the derived $\delta \B$. Consequently, even with an incorrect stellar characterization, our method provides valuable information on the temporal behavior of the magnetic fields, with a damped amplitude.
    Our simulations also underscore the importance of careful relative normalization, as short-window median filters tend to impact the shape of broad absorption features, with significant impacts on both $\delta \B$ and temperature variation estimates ($\delta T$).
    The proposed method allows for the retrieval of extremely fast estimates of the absolute surface magnetic field strength ($\B$) for a series of spectra, provided that the value associated with the reference spectrum is known. Such an initial estimate can be obtained with the methods commonly used to derive field strengths~\citep[e.g.,][]{reiners-2022, cristofari-2023b, kochukhov-2024}.
    
   We implemented a model-driven process following~\citet{artigau-2024} to derive relative temperature variations ($\delta T$) from the observed spectra. For EV~Lac and DS~Leo, we found an excellent agreement between our results and the data-driven $\delta T$ obtained in the framework of the LBL code by~\citep{artigau-2024}. Similarly to~\citet{cristofari-2025b}, we find significant anticorrelation between $\delta T$ and $\delta \B$, suggesting that the observed magnetic fields are directly linked to the presence of cooler spots at the stellar surface.
   Relying on spectra recorded with ESPaDOnS and Narval, we see clear anticorrelations between our retrieved $\delta T$ and $\delta \B$ for EV~Lac and DS~Leo. 
        We obtained similar slopes between  $\delta T$ and $\delta \B$ for the optical and nIR domains, although with a significantly higher scatter in the optical domain.
   Several competing factors could explain these results, including the possibly higher sensitivity to temperature variations in the optical, caused by higher contrast between spots and the photosphere~\citep{berdyugina-2005}. One identified cause, however, could be the influence of magnetic fields on the $\delta T$ estimates. Simulations revealed that magnetic field variations can lead to an increased scatter and larger amplitudes in the derived $\delta T$ (Fig.~\ref{fig:impact-b-on-teff}), particularly for regions containing a large number of magnetically sensitive transitions, which is consistent with the larger scatter observed for ESPaDOnS and Narval data.
   We note that a natural next step could be to rely on the $\delta\B$ measurements to refine $\delta T$ estimates. Using $\delta\B$, we could compute magnetic-field dependent $\partial\vec{A}/\partial T$ (Eq.~\ref{eq:eq_linear_ff}) and used these to obtain more accurate $\delta T$. We leave the detailed implementation of this process to a subsequent study.
   We note that temperature variations can also have a small impact on the derived $\delta \B$, although these appear small in light of the current measurements precision. The next generation of instruments will provide spectra with much higher quality and the implementation of temperature dependent magnetic models, mimicking the impact of cool magnetic spots at the stellar surface, will enable a more detailed modeling of the stellar spectra.
   
   Narval and ESPaDOnS data provide access to the CaII H\&K lines, which we used to measure the S-index, widely used as an activity indicator in Sun-like stars and M dwarfs. We found no clear correlation between the S-index estimates and our $\delta\B$ measurements. These results can largely be attributed to the nature of the activity in M dwarfs, which are more prone to chromospheric activity and flares, potentially  throwing off the relation that typically arises from steady quiescent heating.
        In addition, the relatively low S/N in the blue orders of the ESPaDOnS/Narval spectrograph for M dwarfs can impact our S-index estimates. 
    Observations recorded with these instruments also span close to a decade, while studies looking at the detailed correlation between S-index and $\B$ in the Sun typically focused on much smaller timescales. Complex long-term fluctuations in the magnetic activity could throw off the S-index--$\B$ relation observed in the Sun. These results underscore the need for small-scale magnetic field measurements which are less impacted by chromospheric activity and the need for dense observation campaigns to fully explore the link between surface magnetic fields and activity indicators in low-mass stars.

The current method will benefit from improvement in the models used to obtain the relative variation of pixels with magnetic field strength. The models used in this work were computed under the assumption that only atoms are sensitive to magnetic fields, due to the lack of known Land\'e factors for most magnetically sensitive molecules. While the impact of magnetically sensitive molecules on the results can be mitigated by pixel rejection strategies and avoiding dense molecular bands, including the impact of magnetic fields on sensitive molecules (such as TiO) will allow for the precision of the proposed method to be improved.

   Finally, we find that the proposed method can have significant benefits for the correction of radial velocity curves in the context of exoplanet searches. Recent studies have suggested that $\B$ can be an excellent indicator of activity jitters in radial velocity curves~\citep{haywood-2016, haywood-2022}. The ability to obtain fast and reliable constraints from the same data used to measure RVs opens the door to new strategies for activity jitter correction for extracting planetary signals from the recorded data.

\begin{acknowledgements}
      This project has received funding from the European Research Council (ERC) under the European Union's Horizon 2020 research and innovation programme (grant agreement No 817540, ASTROFLOW).
      This project has received funding from the Dutch Research Council (NWO), with project numbers VI.C.232.041 of the Talent Programme Vici and OCENW.M.22.215 of the research programme "Open Competition Domain Science - M".
      S.H.S. gratefully acknowledges support from NASA EPRV grant 80NSSC21K1037, NASA XRP grant 80NSSC21K0607, and NASA Heliophysics LWS grant NNX16AB79G.
\end{acknowledgements}

\bibliographystyle{aa} 
\bibliography{cfa} 

\begin{appendix}

        \section{Data tables}
        
        Tables~\ref{tab:gl873-data},~\ref{tab:gl410-data}, and~\ref{tab:gl699-data} report our $\delta \B$ and $\delta T$ measurements obtained from spectra recorded with SPIRou for EV~Lac, DS~Leo and Barnard's~star, respectively. Tables~\ref{tab:gl873-data-narval} and~\ref{tab:gl410-data-narval} present similar data obtained from spectra recorded with Narval/ESPaDOnS for EV~Lac and DS~Leo, respectively.
        
\begin{table}
        \caption{$\delta \B$ and $\delta T$ measurements obtained for EV~Lac from spectra recorded with SPIRou.\label{tab:gl873-data}}
        \begin{tabular}{ccc}
                \hline
                \hline
                MJD & $\langle B \rangle$ (kG) & $\delta T$ (K) \\
                \hline
58737.5820 & $0.03\pm0.05$ & $-0.71\pm0.85$\\
58744.3863 & $-0.07\pm0.06$ & $1.38\pm0.85$\\
58745.3477 & $-0.32\pm0.06$ & $0.28\pm0.87$\\
                \hline
        \end{tabular}
        \tablefoot{The complete table is available at the CDS in machine readable format.}
\end{table}
        
        \begin{table}
                \caption{Same as Table~\ref{tab:gl873-data} but for DS~Leo.\label{tab:gl410-data}}
                \begin{tabular}{ccc}
                        \hline
                        \hline
                        MJD & $\delta\langle B \rangle$ (kG) & $\delta T$ (K) \\
                        \hline
59157.6134 & $0.20\pm0.03$ & $-7.72\pm0.81$\\
59158.6077 & $0.21\pm0.03$ & $-12.72\pm0.80$\\
59207.5509 & $-0.05\pm0.03$ & $-0.87\pm0.82$\\
                \hline
                \end{tabular}
\end{table}

        \begin{table}
        \caption{Same as Table~\ref{tab:gl873-data}, but for Barnard's star.\label{tab:gl699-data}}
        \begin{tabular}{ccc}
                \hline
                \hline
                MJD & $\delta\langle B \rangle$ (kG) & $\delta T$ (K) \\
                \hline
58528.6717 & $0.18\pm0.04$ & $-4.22\pm0.63$\\
58530.6699 & $0.07\pm0.04$ & $-5.16\pm0.62$\\
58531.6855 & $-0.12\pm0.05$ & $-0.72\pm0.89$\\
                \hline
        \end{tabular}
\end{table}

\begin{table}
        \caption{$\delta \B$ and $\delta T$ measurements obtained for EV~Lac from spectra recorded with Narval/ESPaDOnS (extract).\label{tab:gl873-data-narval}}
        \begin{tabular}{ccc}
                \hline
                \hline
                MJD & $\langle B \rangle$ (kG) & $\delta T$ (K) \\
                \hline
53568.6300 & $0.07\pm0.03$ & $1.20\pm0.48$\\
53605.6471 & $0.24\pm0.03$ & $-0.56\pm0.61$\\
53631.4915 & $0.21\pm0.03$ & $-1.80\pm0.46$\\
                \hline
        \end{tabular}
        \tablefoot{The complete table is available at the CDS.}
\end{table}

        \begin{table}
        \caption{Same as Table~\ref{tab:gl873-data-narval} but for DS~Leo.\label{tab:gl410-data-narval}}
        \begin{tabular}{ccc}
                \hline
                \hline
                MJD & $\delta\langle B \rangle$ (kG) & $\delta T$ (K) \\
                \hline
53747.6057 & $0.22\pm0.02$ & $3.98\pm0.28$\\
53773.5379 & $0.20\pm0.02$ & $2.01\pm0.28$\\
54127.1466 & $-0.06\pm0.02$ & $-0.48\pm0.32$\\
                \hline
        \end{tabular}
\end{table}
        
        \section{GP fit parameters}
        
        Table~\ref{tab:gp-spirou} presents the GP parameters retrieved from posterior distribution by applying our process to the spectra recorded with SPIRou. Table~\ref{tab:gp-narval} presents the same parameters obtained by applying our process to the spectra recorded with Narval/ESPaDOnS.
        
                \begin{table}
                        \renewcommand{\arraystretch}{1.25}
                        \caption{GP hyper-parameters obtained for EV~Lac, DS~Leo, and Barnard's star from SPIRou spectra.}
                        \label{tab:gp-spirou}
                        \begin{tabular}{lcc}
                                \hline
                                \hline 
                                & Value & Prior \\ 
                                \hline 
                                \multicolumn{3}{c}{GP fit for EV Lac}\\$\mu$~(kG) & $-0.10^{+0.19}_{-0.20}$ & $\mathcal{U}(-10, 10)$ \\ 
                                $\sigma$~(kG) & $0.03^{+0.01}_{-0.01}$ & $\mathcal{U}(0, 1000)$ \\ 
                                $\alpha$~(kG) & $0.33^{+0.14}_{-0.09}$ & $\mathcal{U}(0, 100)$ \\ 
                                $l$~(d) & $420^{+73}_{-64}$ & $\mathcal{U}(100, 800)$ \\ 
                                $\beta$ & $0.60^{+0.13}_{-0.11}$ & $\mathcal{U}(0, 10)$ \\ 
                                $P_{\rm rot}$~(d) & $4.361^{+0.001}_{-0.001}$ & $\mathcal{U}(0, 10)$ \\ 
                                $\chi^2_r$ & 1.05 &  \\
                                RMS & $0.06$~(kG) &  \\
                                \hline
                                \multicolumn{3}{c}{GP fit for DS~Leo}\\$\mu$~(kG) & $0.01^{+0.02}_{-0.02}$ & $\mathcal{U}(-5, 5)$ \\ 
                                $\sigma$~(kG) & $0.01^{+0.00}_{-0.01}$ & $\mathcal{U}(0, 1000)$ \\ 
                                $\alpha$~(kG) & $0.08^{+0.01}_{-0.01}$ & $\mathcal{U}(0, 10)$ \\ 
                                $l$~(d) & $58^{+7}_{-7}$ & $\mathcal{U}(30, 500)$ \\ 
                                $\beta$ & $0.61^{+0.10}_{-0.09}$ & $\mathcal{U}(0, 20)$ \\ 
                                $P_{\rm rot}$~(d) & $14.063^{+0.080}_{-0.081}$ & $\mathcal{U}(5, 20)$ \\ 
                                $\chi^2_r$ & 0.74 &  \\
                                RMS & $0.02$~(kG) &  \\
                                \hline 
                                \multicolumn{3}{c}{GP fit for Barnard's~star}\\$\mu$~(kG) & $0.02^{+0.03}_{-0.03}$ & $\mathcal{U}(-10, 10)$ \\ 
                                $\sigma$~(kG) & $0.00^{+0.00}_{-0.00}$ & $\mathcal{U}(0, 1000)$ \\ 
                                $\alpha$~(kG) & $0.11^{+0.02}_{-0.02}$ & $\mathcal{U}(0, 10)$ \\ 
                                $l$~(d) & $137^{+26}_{-29}$ & $\mathcal{U}(50, 500)$ \\ 
                                $\beta$ & $0.60^{+0.12}_{-0.10}$ & $\mathcal{U}(0, 20)$ \\ 
                                $P_{\rm rot}$~(d) & $146.338^{+7.961}_{-5.333}$ & $\mathcal{U}(50, 200)$ \\ 
                                $\chi^2_r$ & 0.82 &  \\
                                RMS & $0.05$~(kG) &  \\
                                \hline
                        \end{tabular}
                        \tablefoot{The hyper-parameters of the kernel are the mean value ($\mu$), the standard deviation of the added uncorrelated noise ($\sigma$), the amplidute ($\alpha$), the decay time ($l$), the smoothing factor $\beta$ and the recurrence period ($P_{\rm rot}$).}
                \end{table}

                \begin{table}
                        \renewcommand{\arraystretch}{1.25}
                        \caption{GP hyper-parameters obtained for EV~Lac and DS~Leo from Narval/ESPaDOnS spectra.}
                        \label{tab:gp-narval}
                        \begin{tabular}{lcc}
                                \hline
                                \hline 
                                        & Value & Prior \\ 
                                        \hline 
                                        \multicolumn{3}{c}{GP fit for EV~Lac }\\
$\mu$~(kG) & $0.02^{+0.06}_{-0.06}$ & $\mathcal{U}(-10, 10)$ \\ 
$\sigma$~(kG) & $0.08^{+0.01}_{-0.01}$ & $\mathcal{U}(0, 1000)$ \\ 
$\alpha$~(kG) & $0.19^{+0.05}_{-0.03}$ & $\mathcal{U}(0, 100)$ \\ 
$l$~(d) & $266^{+91}_{-87}$ & $\mathcal{U}(100, 800)$ \\ 
$\beta$ & $0.44^{+0.16}_{-0.14}$ & $\mathcal{U}(0, 10)$ \\ 
$P_{\rm rot}$~(d) & $4.367^{+0.023}_{-0.011}$ & $\mathcal{U}(0, 10)$ \\ 
$\chi^2_r$ & 3.84 &  \\
RMS & $0.06$~(kG) &  \\
                                        \hline 
                                        \multicolumn{3}{c}{GP fit for DS~Leo}\\
                                        $\mu$~(kG) & $0.05^{+0.04}_{-0.04}$ & $\mathcal{U}(-5, 5)$ \\ 
                                        $\sigma$~(kG) & $0.00^{+0.00}_{-0.00}$ & $\mathcal{U}(0, 1000)$ \\ 
                                        $\alpha$~(kG) & $0.10^{+0.04}_{-0.02}$ & $\mathcal{U}(0, 10)$ \\ 
                                        $l$~(d) & $80^{+29}_{-23}$ & $\mathcal{U}(30, 500)$ \\ 
                                        $\beta$ & $0.79^{+0.83}_{-0.58}$ & $\mathcal{U}(0, 20)$ \\ 
                                        $P_{\rm rot}$~(d) & $14.109^{+0.406}_{-0.421}$ & $\mathcal{U}(5, 20)$ \\ 
                                        $\chi^2_r$ & 0.36 &  \\
                                        RMS & $0.01$~(kG) &  \\
                                        \hline
                        \end{tabular}
                \end{table}

        \section{Additional comparisons}
                
\begin{figure}
                        \centering
                        \includegraphics[width=0.80\linewidth]{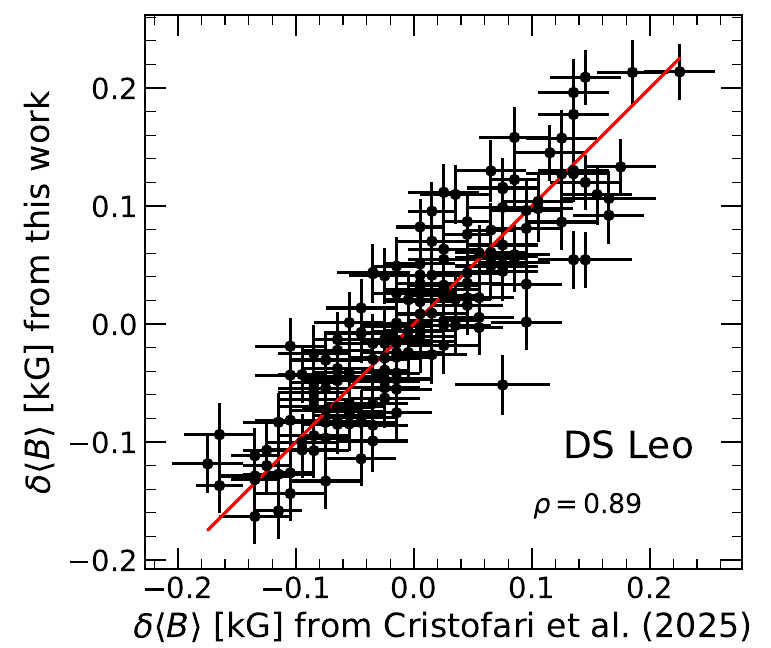}\vspace{5pt}
                        \includegraphics[width=0.80\linewidth]{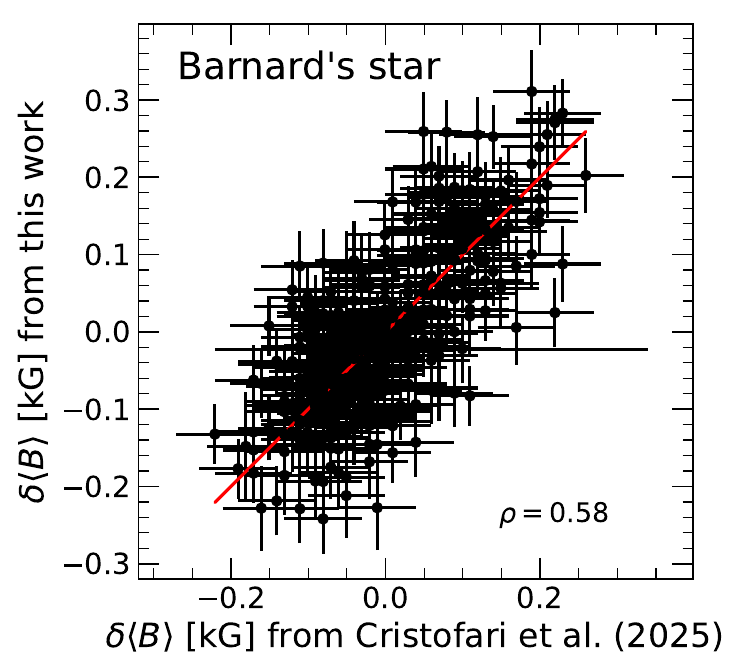}
                        \caption{Same as Fig.~\ref{fig:compare-gl873}, but for DS~Leo (top panel) and Barnard's star (bottom panel).}
                        \label{fig:compare-gl410+gl699}
                \end{figure}

\begin{figure}[h!]
                        \centering
                        \includegraphics[width=0.80\linewidth]{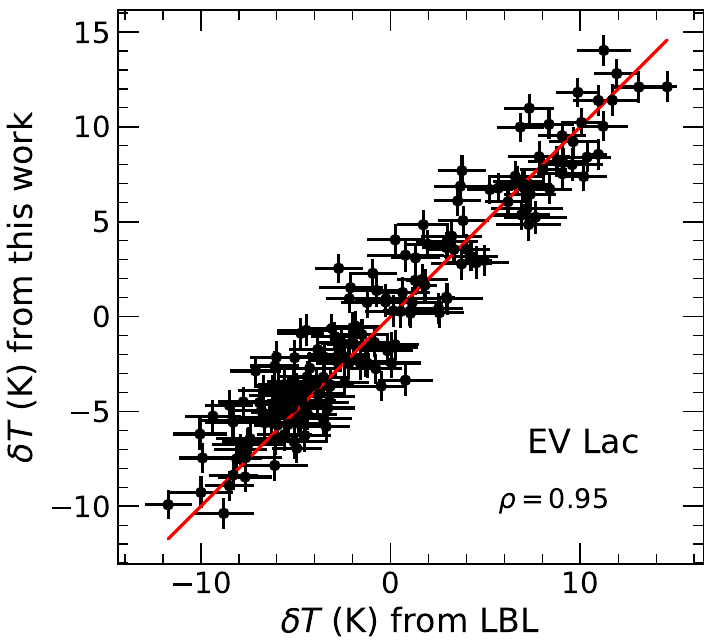}\vspace{10pt}
                        \includegraphics[width=0.80\linewidth]{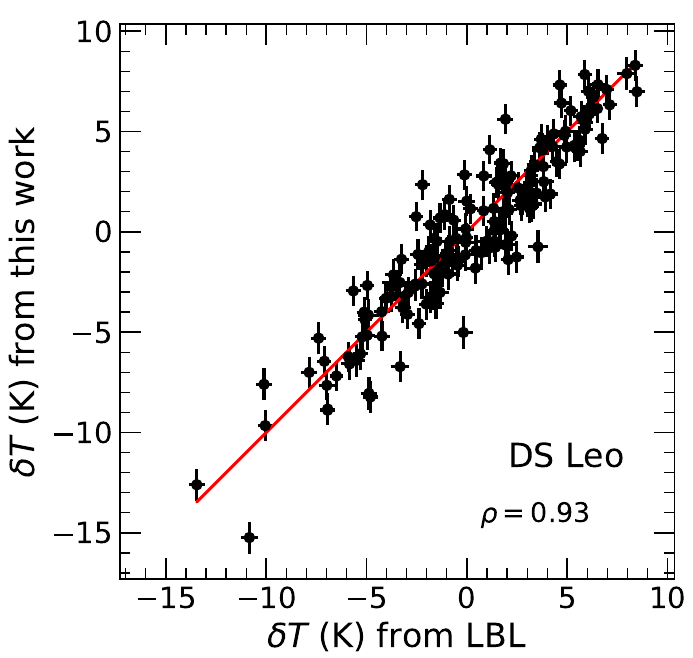}\vspace*{10pt}
                        \includegraphics[width=0.80\linewidth]{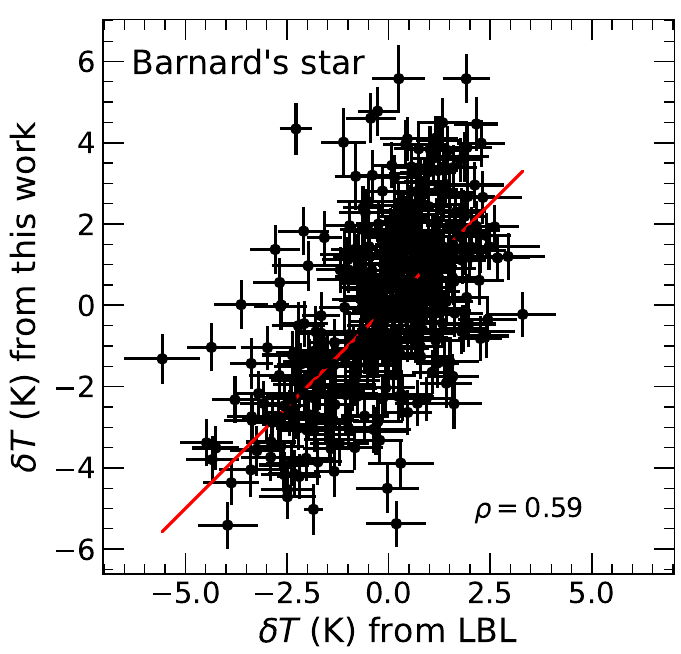}
                        \caption{Comparison between our model-driven $\delta T$ and the data-driven estimates obtained with the LBL~\citep[see][]{artigau-2024,cristofari-2025b} for EV~Lac (top panel), DS~Leo (middle panel), and Barnard's~star (bottom panel).
                                The red line shows the equality.}
                        \label{fig:compare-dtemp-gl873+gl410+gl699}
                \end{figure}

                Figure~\ref{fig:compare-gl410+gl699} 
                presents the comparison between $\delta\B$ obtained in this paper, and those of~\citet{cristofari-2025b}. Figure~\ref{fig:compare-dtemp-gl873+gl410+gl699}
                presents comparisons between the retrieved $\delta T$ and the model-driven variations obtained with the \texttt{LBL}.
                
                \section{$\delta \B$ -- $\delta T$ relations}
                
\begin{figure}
                        \includegraphics[width=\linewidth]{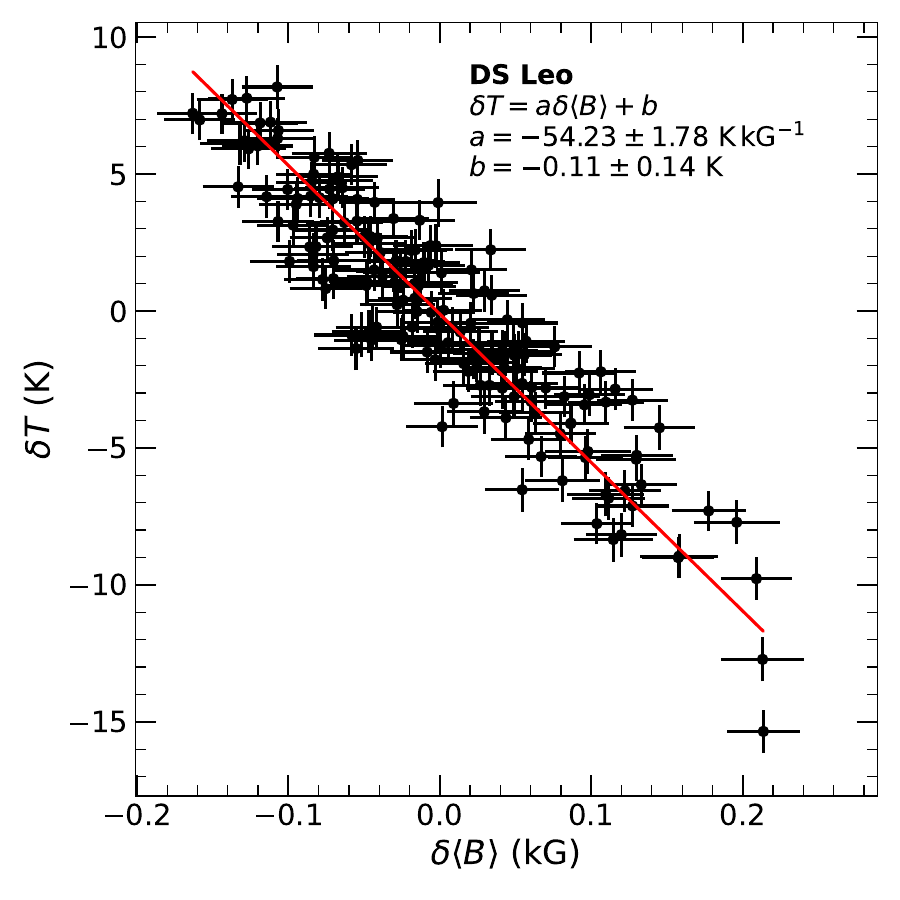}
                        \includegraphics[width=\linewidth]{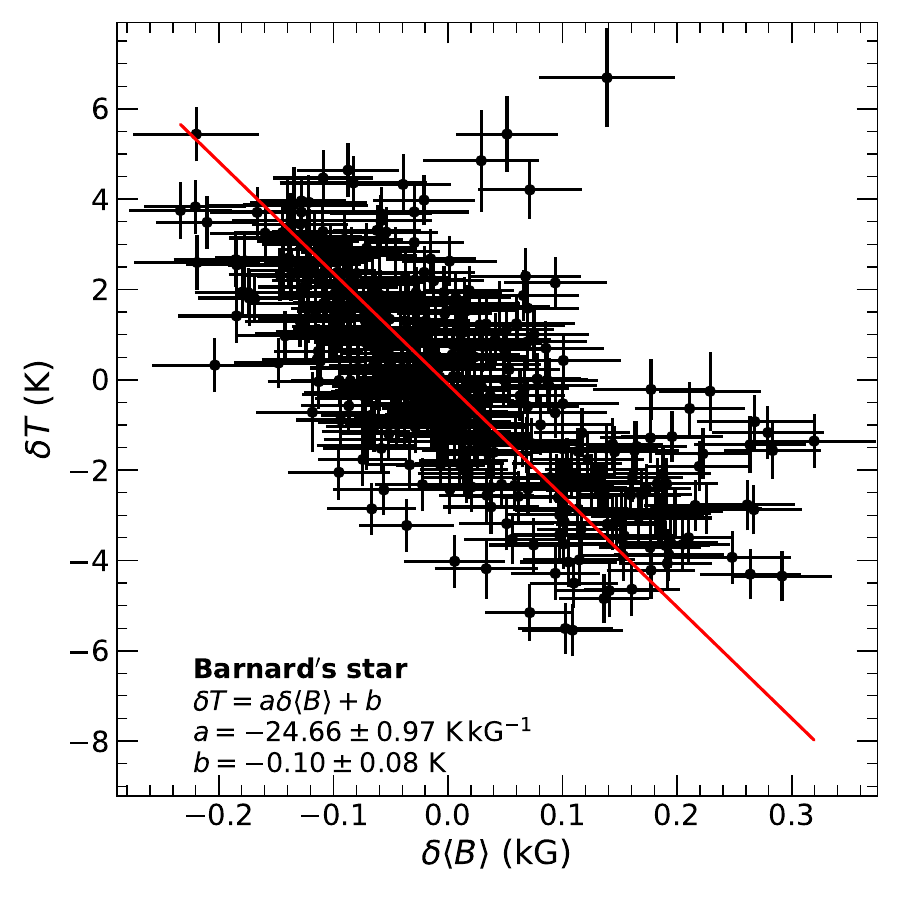}
                        \caption{Same as Fig.~\ref{fig:correlation_teff_b_spirou_gl873} but for DS~Leo (top panel) and Barnard's star (bottom panel).}
                        \label{fig:correlation_teff_b_spirou_gl410+gl699}
                \end{figure}

\begin{figure}
                        \includegraphics[width=\linewidth]{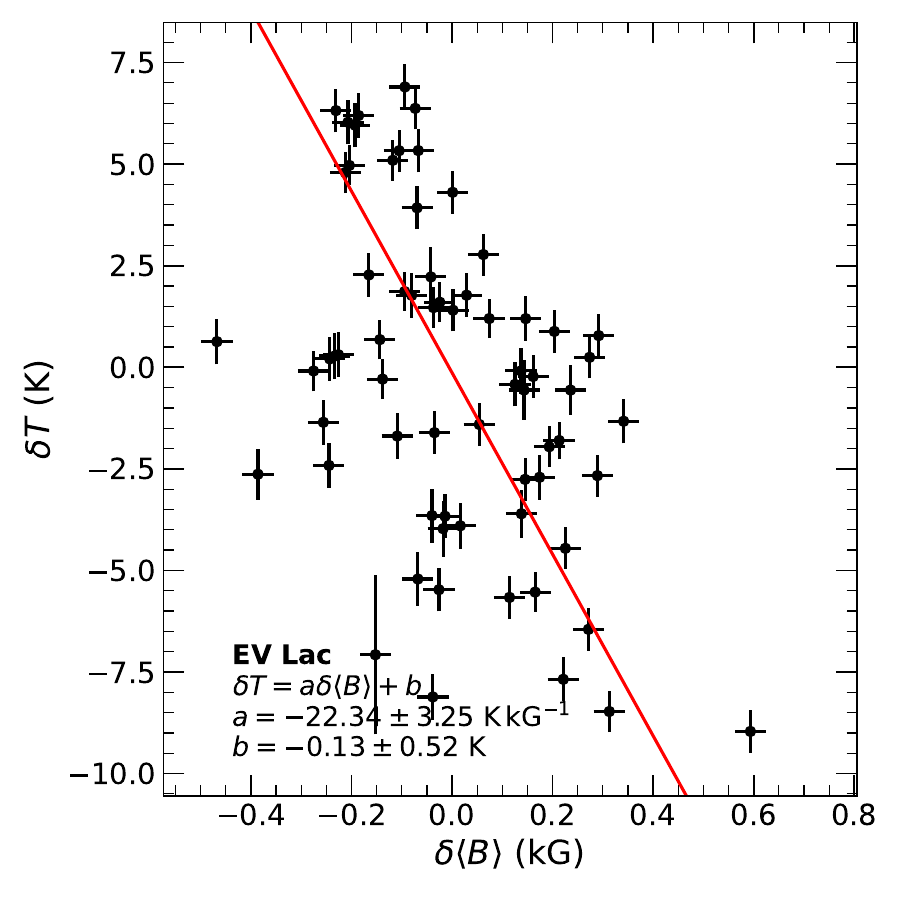}
                        \includegraphics[width=\linewidth]{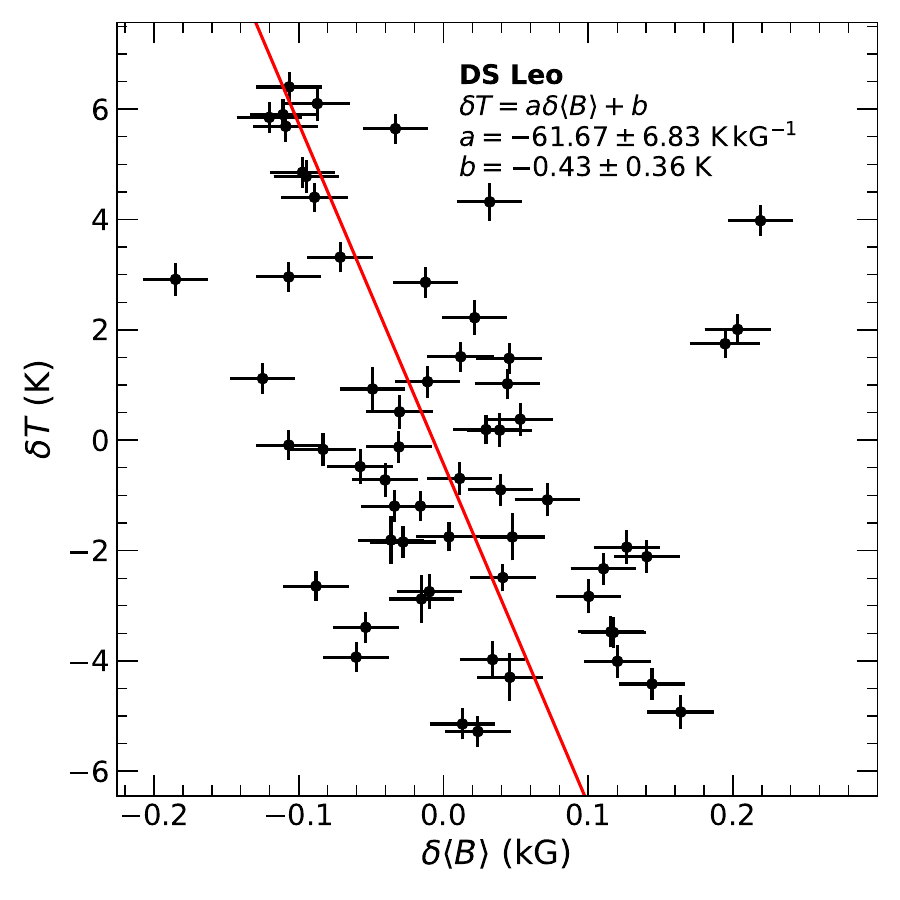}
                        \caption{Relation between our model $\delta T$ and $\delta \B$ estimates obtained from Narval/ESPaDOnS data for EV~Lac (top panel) and DS~Leo (bottom panel). The red line shows the best linear fits obtained with an orthogonal distance regression. The parameters of the fits are shown on the figure.}
                        \label{fig:correlation_teff_b_narval_gl873+gl410}
                \end{figure}

                Figure~\ref{fig:correlation_teff_b_spirou_gl410+gl699}  
                presents the comparison between our derived $\delta\B$ and model-driven $\delta T$ obtained from SPIRou spectra for DS~Leo and Barnard's~star. Figure~\ref{fig:correlation_teff_b_narval_gl873+gl410} 
                presents similar comparisons for the estimates obtained from Narval/ESPaDOnS data, for EV~Lac and DS~Leo.

        \section{Corner plots}
\begin{figure}
                \includegraphics[width=\linewidth]{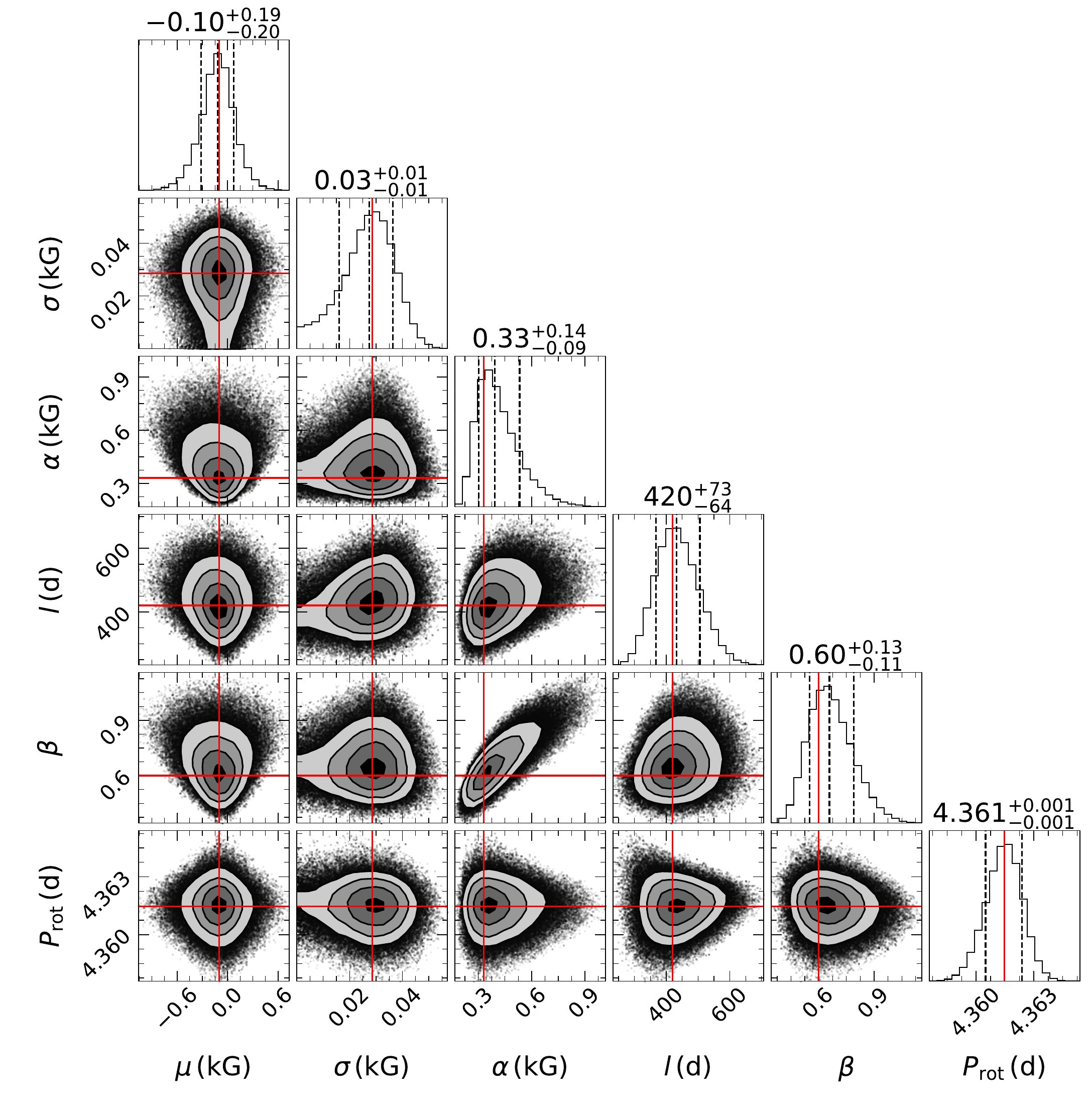}
                \caption{Posterior distribution walkers for the GP parameters obtained on $\delta\B$ extracted from the spectra recorded with SPIRou for EV~Lac.}
                \label{fig:spirou-gl873-corner}
        \end{figure}

\begin{figure}
                \includegraphics[width=\linewidth]{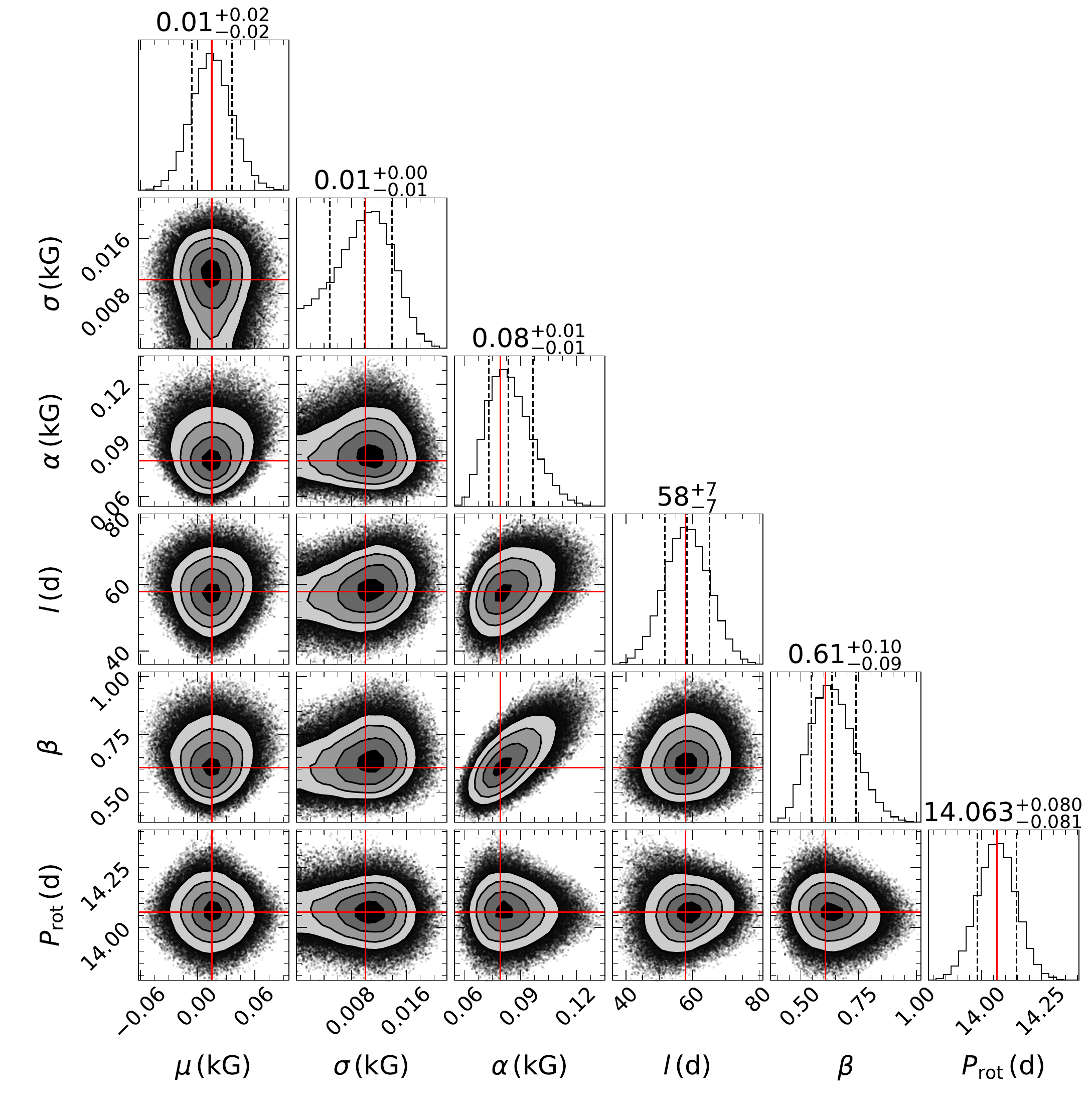}
                \caption{Same as Fig.~\ref{fig:spirou-gl873-corner}, but for DS~Leo.}
                \label{fig:spirou-gl410-corner}
        \end{figure}

\begin{figure}
                \includegraphics[width=\linewidth]{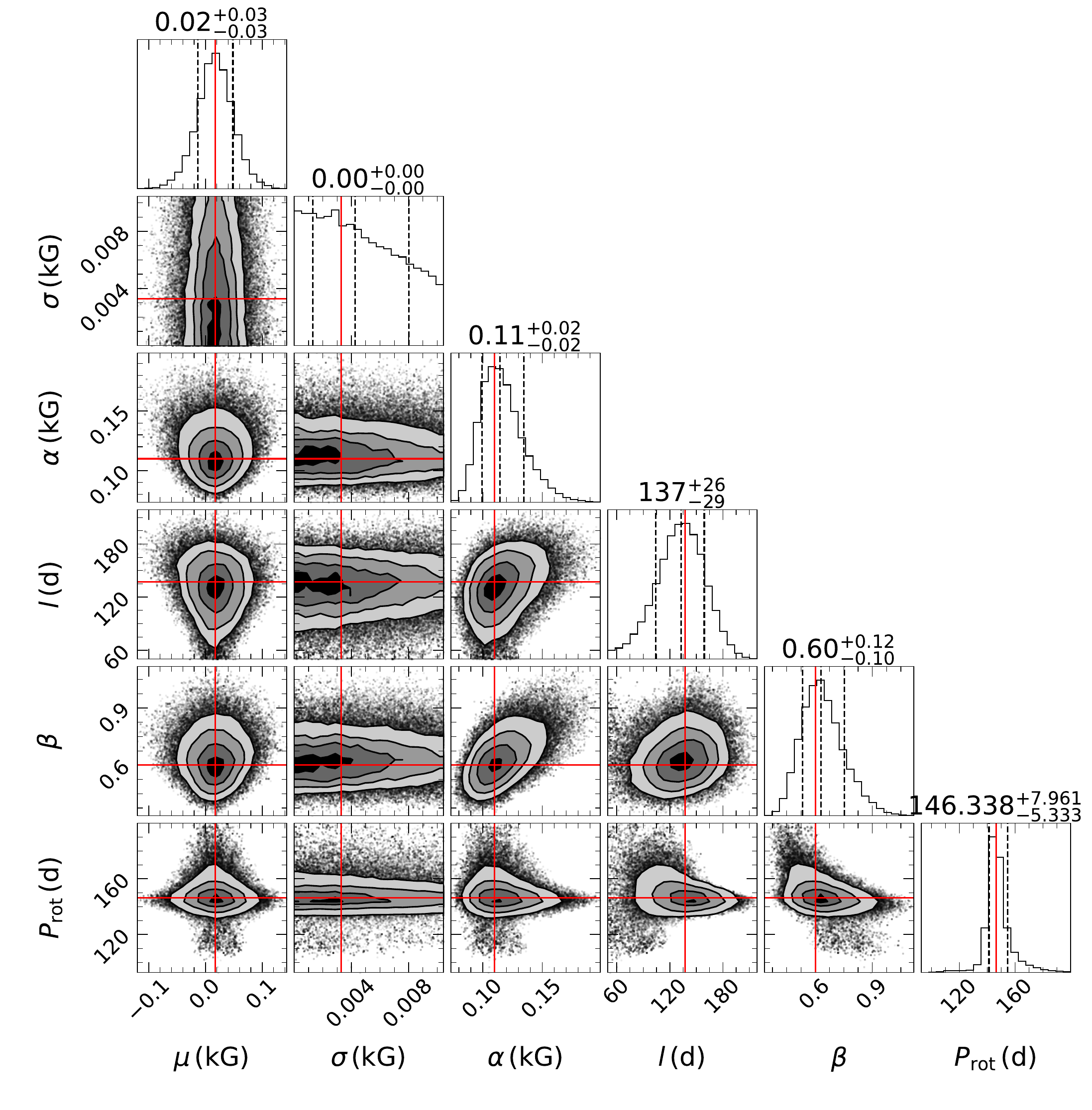}
                \caption{Same as Fig.~\ref{fig:spirou-gl873-corner}, but for Barnard's~star.}
                                \label{fig:spirou-gl699-corner}
        \end{figure}

\begin{figure}
                \includegraphics[width=\linewidth]{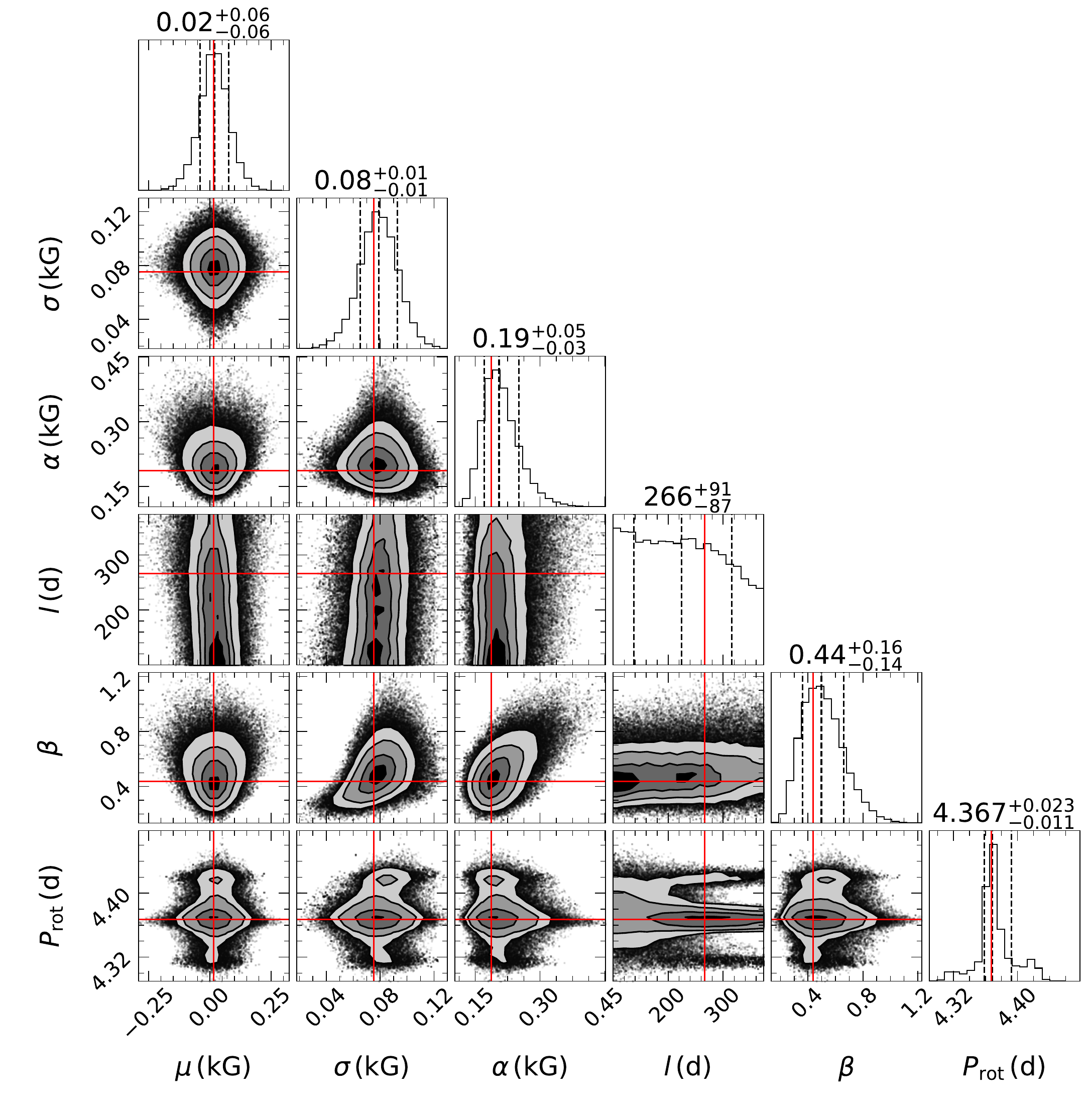}
                \caption{Same as Fig.~\ref{fig:spirou-gl873-corner}, but for EV~Lac from data recorded with Narval/ESPaDOnS.}
                \label{fig:narval-gl873-corner}
        \end{figure}

\begin{figure}
                \includegraphics[width=\linewidth]{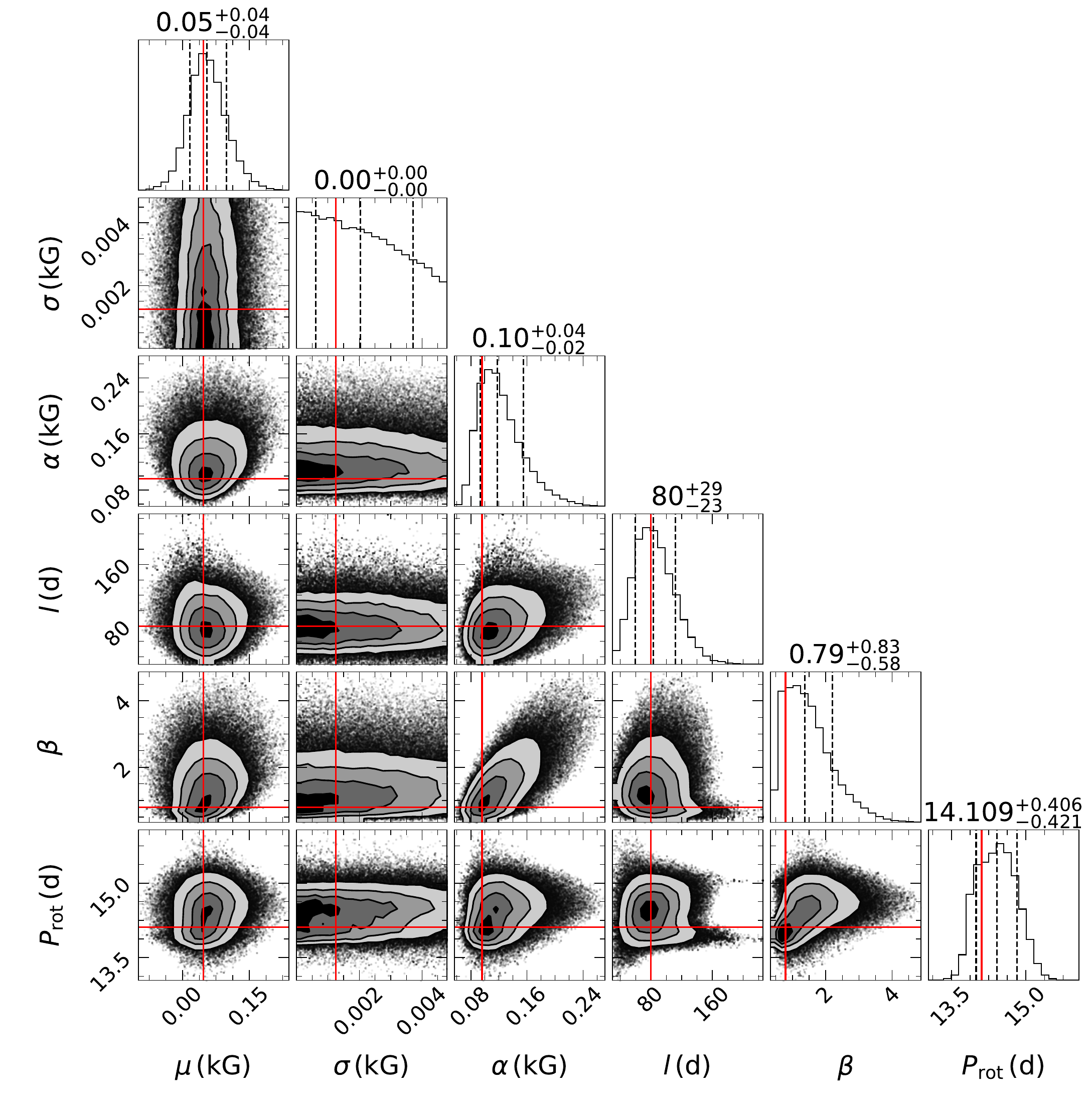}
                \caption{Same as Fig.~\ref{fig:spirou-gl873-corner}, but for DS~Leo from data recorded with Narval/ESPaDOnS.}
                \label{fig:narval-gl410-corner}
        \end{figure}

        Figures~\ref{fig:spirou-gl873-corner},~\ref{fig:spirou-gl410-corner}, and~\ref{fig:spirou-gl699-corner} present the corner plots showing the posterior distributions of the hyper-parameters associated with the quasi-periodic Gaussian Fit on $\delta\B$ estimates obtained from SPIRou spectra for EV~Lac, DS~Leo and Barnard's star, respectively. Figures~\ref{fig:narval-gl873-corner} and~\ref{fig:narval-gl410-corner} present similar corner plots obtained with $\delta\B$ estimates derived from Narval/ESPaDOnS spectra for EV~Lac and DS~Leo, respectively.

\section{S-index calculation}
\begin{figure}
        \includegraphics[width=2\linewidth]{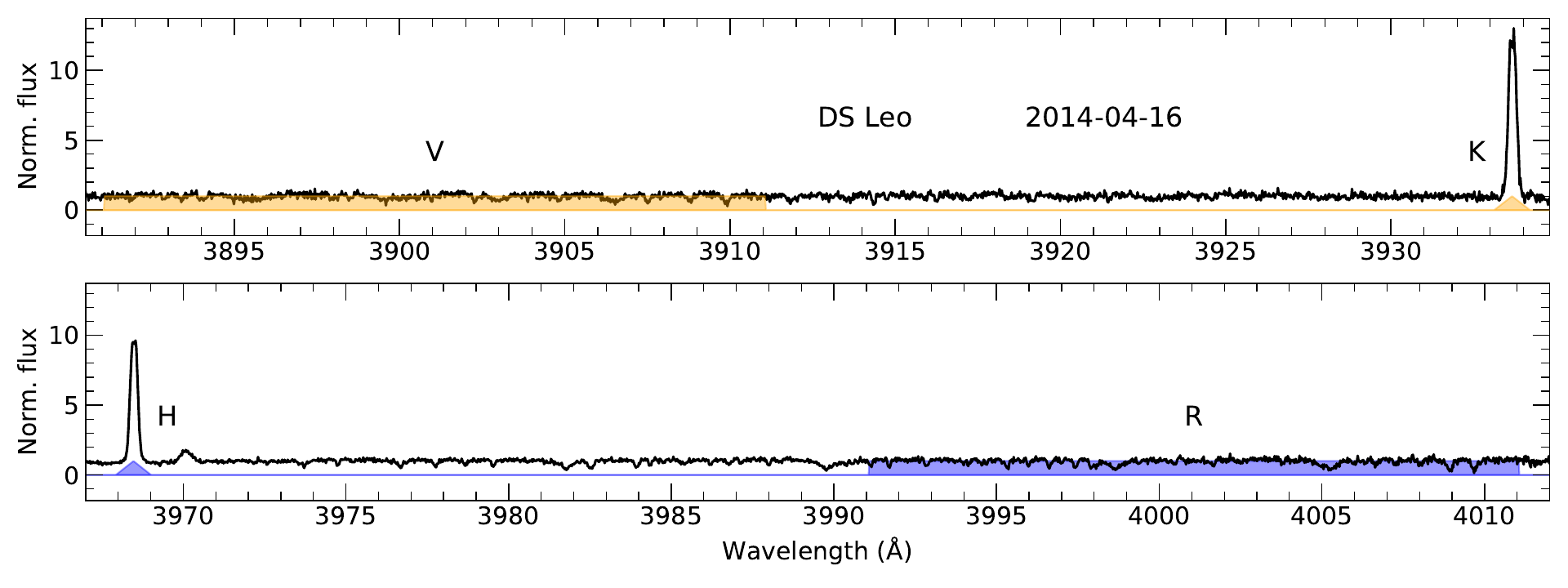}
        \caption{Example of spectrum obtained with Narval, showing the V and R rectangular bands, along with the K and H triangular bands used to compute the S-index.}
        \label{fig:example_bands_sindex}
\end{figure}

Figure~\ref{fig:example_bands_sindex} shows an example of spectrum showing the bands used to compute the S-index.

\end{appendix}

\end{document}